\documentclass{article}

\PassOptionsToPackage{numbers,compress}{natbib}
\usepackage[main, preprint]{neurips_2026}

\usepackage[utf8]{inputenc} 
\usepackage[T1]{fontenc}    
\usepackage{hyperref}       
\usepackage{url}            
\usepackage{booktabs}       
\usepackage{amsfonts}       
\usepackage{amssymb}
\usepackage{amsmath}
\usepackage{nicefrac}       
\usepackage{microtype}      
\usepackage{xcolor}         
\usepackage{graphicx}
\usepackage{algorithm}
\usepackage{algorithmic}
\usepackage{subcaption}
\usepackage{multirow}
\usepackage{listings}
\graphicspath{ {./images/} }

\title{Noise-Regularized Training for Learned Image Compression}

\author{%
  Renjie Zou \quad Zhiwei Huang\thanks{Corresponding authors: Zhiwei Huang and Dong Jiang.} \quad Dong Jiang\footnotemark[1]\\
  Malanshan Audio \& Video Laboratory\\
  \texttt{\{zourenjie,huangzhiwei,jiangdong\}@mlslabs.com.cn}
}

\begin{document}

\maketitle

\begin{abstract}
Learned image compression (LIC) is trained almost exclusively on
clean images: the only stochastic perturbation in the loop is an
additive uniform proxy injected in the \emph{latent} domain to keep
the entropy model differentiable. Outside compression, by contrast,
a long line of work---from Tikhonov regularization and contractive
autoencoders through denoising score matching and modern diffusion
models---has established that Gaussian noise at the \emph{input} of
a network with a clean target acts as a structured probe of the
data manifold rather than as data augmentation. Building on this
view, we introduce \emph{noise-regularized training} for LIC: at
every training step we add Gaussian noise to the encoder input
while keeping the distortion target, the latent uniform proxy, the
hyperprior, and the entropy coding pipeline unchanged. The recipe
is architecture-agnostic and adds essentially no training or
inference overhead. Across backbones and operating regimes it
yields consistent BD-Rate gains: $-2.60\%$ on a reproduced ELIC and
$-0.85\%/-1.05\%/-2.38\%$ on the Base/Medium/Large scales of a
variable-rate DMCI codec, with DMCI-Large reaching $-22.02\%$
against VTM-22.0. A second-order expansion of the noisy RD loss
accounts for these gains: because the noise enters \emph{upstream}
of the analysis transform, it converts the clean RD objective into
a \emph{structured} $O(\sigma^{2})$ regularizer comprising an
information-weighted encoder-Jacobian penalty (rate side) and a
Frobenius contractive penalty on the end-to-end reconstruction map
(distortion side). A finite-difference Lipschitz analysis shows
corresponding sensitivity reductions of $\sim\!12\%$ for the encoder
and $\sim\!8\%$ for the end-to-end codec, while an exact-JVP test
further shows preferential contraction of high-information directions.
Code and trained models are available at
\url{https://github.com/Googolxx/NoiseRegu-LIC}.
\end{abstract}

\section{Introduction}
\label{sec:introduction}

Learned image compression (LIC)~\cite{balle2017end,balle2018variational,minnen2018joint,cheng2020learned,he2021checkerboard,zou2022devil,liu2023tcm,jiang2023mlic}
now rivals and surpasses
VVC~\cite{vvc} on standard benchmarks~\cite{kodak} by jointly optimizing
an analysis transform $g_a$, a synthesis transform $g_s$, and a
differentiable entropy model under the rate--distortion (RD) objective
$\mathcal{L} = R + \lambda\,D$. Effort at the field level has
concentrated on two axes: more expressive nonlinear
transforms~\cite{cheng2020learned,zou2022devil,liu2023tcm,li2024llic} and
more accurate entropy models~\cite{minnen2018joint,he2021checkerboard,minnen2020channel,jiang2023mlic,kim2022joint,qian2022entroformer}.
The training signal itself, however, has remained largely fixed: the
codec is fed clean images and supervised against the same clean target,
and the only stochastic perturbation in the loop is an additive
\emph{uniform} proxy injected in the \emph{latent} domain to keep the
entropy model differentiable~\cite{balle2017end}.

\paragraph{A different role for noise.}
Outside compression, a parallel literature has long taught a
complementary lesson. Bishop~\cite{bishop1995training} showed that
Gaussian input noise is equivalent to Tikhonov regularization, and
Rifai et al.~\cite{rifai2011contractive} used the same Jacobian-Frobenius
penalty to derive contractive autoencoders. More importantly,
Vincent~\cite{vincent2011connection} and Alain \&
Bengio~\cite{alain2014regularized} proved that a denoising autoencoder
trained with Gaussian input corruption and a clean target implicitly
learns the Stein score $\nabla\log p_\sigma$ of the noise-smoothed data
distribution; modern score-based generative models~\cite{song2019generative,song2021sde,karras2022edm}
turn this into a generative procedure by walking samples \emph{onto}
the data manifold via Langevin dynamics across a noise-scale ladder.
The shared takeaway is that noise injected at the \emph{input} of a
network with a clean target is not augmentation but a structured
probe of the geometry of the data distribution.

\paragraph{From latent noise to input noise.}
The latent-domain proxy used in standard LIC is, by contrast, an
\emph{operator approximation}: it mimics the statistics of rounding
but never asks the encoder to be insensitive to plausible variations
in the input image. Inspired by the score matching view above, we
study the simplest pixel-domain counterpart of that recipe: at every
training step we replace the encoder input by
$\tilde{\mathbf{x}} = \mathbf{x} + \boldsymbol{\eta}$ with
$\boldsymbol{\eta}\!\sim\!\mathcal{N}(0,\sigma^{2}I)$, while leaving
the distortion target, the latent uniform proxy, the hyperprior, and
the entropy coding pipeline all untouched. There is no architectural
change and essentially no overhead at either end of the pipeline:
training adds a single additive perturbation in data preprocessing,
and at test time the codec runs unmodified.

\paragraph{Empirical effectiveness.}
This single change yields consistent gains across architectures and bitrates, achieving a $-2.60\%$ BD-Rate on a reproduced ELIC~\cite{he2022elic}. It also improves every scale of the variable-rate DMCI (the intra-image codec of DCVC-RT~\cite{jia2025dcvcrt}), with the strongest variant reaching $-22.02\%$ against VTM-22.0 on Kodak~\cite{kodak}, Tecnick~\cite{tecnick}, and CLIC~\cite{clic} (Table~\ref{tab:bdrate_results}). These results motivate the central
question of the paper: what regularization does pixel-domain input
noise impose on the RD objective, and why is it consistently
beneficial?

\paragraph{Theoretical picture.}
A second-order expansion of the noisy RD loss
(Section~\ref{sec:theory}) shows that, because the noise enters
\emph{upstream} of $g_a$, it adds two structured $O(\sigma^{2})$
corrections: a Frobenius contractive penalty on the end-to-end
Jacobian (LIC analogue of the contractive
autoencoder~\cite{rifai2011contractive}) and an
\emph{information-weighted} encoder-Jacobian penalty rescaled by the
entropy model's curvature, so directions expensive to encode are
penalized most. Empirically, the encoder's local Lipschitz drops by
$\sim\!12\%$ and the end-to-end codec's by $\sim\!8\%$, while an
exact-JVP test further confirms stronger contraction of
high-$I_\phi$ directions
(Fig.~\ref{fig:lipschitz}, Tables~\ref{tab:lipschitz_summary} and
\ref{tab:info_weighting}).

\paragraph{Contributions.}
\begin{itemize}
  \item \textbf{A pixel-domain noise-regularization recipe for LIC.}
    Motivated by the score matching view of input noise as a manifold
    probe, we \emph{add} a zero-mean Gaussian perturbation at the
    encoder input while leaving the standard latent uniform proxy and
    everything else in the pipeline unchanged, with a clean distortion
    target. The change is architecture-agnostic and adds essentially
    no training or inference overhead.
  \item \textbf{A regularization-based explanation.} A local
    second-order expansion of the noisy RD loss shows that the
    resulting training signal is a \emph{structured} regularizer: an
    information-weighted encoder Jacobian penalty (rate side) plus a
    Frobenius contractive penalty on the end-to-end reconstruction map
    (distortion side), with a closed-form LIC-specific residual--Hessian
    correction that is forced to survive by the rate constraint.
  \item \textbf{State-of-the-art RD performance and direct geometric
    evidence.} The recipe improves a reproduced ELIC by
    $-2.60\%$ BD-Rate on average and every DMCI scale by
    $-0.85\%/-1.05\%/-2.38\%$ (Base/Medium/Large); DMCI-Large reaches
    $-22.02\%$ against VTM-22.0 without any architectural modification.
    A finite-difference Lipschitz analysis shows both predicted
    contractions ($\sim\!12\%$ encoder, $\sim\!8\%$ end-to-end
    codec), and an exact-JVP test shows preferential contraction of
    high-$I_\phi$ directions.
\end{itemize}

\section{Related Work}

\paragraph{Learned image compression.}
LIC has progressed along two main axes. \emph{Nonlinear transforms} have
evolved from GDN-based convolutions~\cite{balle2017end,balle2018variational}
through attention~\cite{cheng2020learned}, window-based
Transformers~\cite{zou2022devil,zhu2022transformerbased}, hybrid
CNN--Transformer backbones~\cite{liu2023tcm}, large-receptive-field and
adaptive transforms~\cite{li2024llic}, and
linear-attention~\cite{feng2025lalic}. \emph{Entropy models} have similarly
advanced from scale hyperpriors~\cite{balle2018variational} through
autoregressive~\cite{minnen2018joint}, checkerboard~\cite{he2021checkerboard},
channel-wise~\cite{minnen2020channel}, global--local
attention~\cite{kim2022joint}, multi-reference~\cite{jiang2023mlic},
progressive context~\cite{li2025hpcm}, and
Transformer-based~\cite{qian2022entroformer} context models. All of
these improvements optimize a \emph{clean-input} RD loss; the
training-signal modification we propose is orthogonal and combines
with all of them. A few works examine training dynamics for the
quantization step itself, e.g.\ straight-through estimators
applied to LIC~\cite{theis2017lossy} or variable-rate
conditioning~\cite{choi2019variable,cui2021asymmetric}, but
\emph{input-domain} Gaussian noise has been largely ignored.

\paragraph{Input noise as a manifold probe.}
A separate line of work has long studied input Gaussian noise as a
tool to shape the geometry of a learned representation.
Bishop~\cite{bishop1995training} first showed that it is equivalent
to Tikhonov regularization. Vincent et al.~\cite{vincent2008extracting}
introduced the denoising autoencoder, and
Rifai et al.~\cite{rifai2011contractive} made the connection to
Jacobian-Frobenius penalization explicit. The decisive theoretical
step came from Vincent~\cite{vincent2011connection} and Alain \&
Bengio~\cite{alain2014regularized}: a noisy-input, clean-target
denoising autoencoder trained at corruption level $\sigma$ implicitly
learns the Stein score $\nabla\log p_\sigma$ of the noise-smoothed
data distribution. Score-based generative
models~\cite{song2019generative,ho2020denoising,song2021sde,karras2022edm}
later turned this observation into a generative procedure, walking
samples onto the data manifold via Langevin dynamics along a
noise-scale ladder. The shared takeaway, that input noise paired with
a clean target acts as a structured probe of the data manifold, is
the conceptual basis of our method; to our knowledge it has not
previously been imported into the compression setting.

\paragraph{Latent vs.\ input noise in LIC.}
Standard LIC injects an additive uniform proxy
$\tilde{\mathbf{y}} = \mathbf{y} + \mathbf{u}$,
$\mathbf{u} \sim \mathcal{U}(-\tfrac{1}{2}, \tfrac{1}{2})$, in the
\emph{latent} domain purely as a continuous relaxation of
quantization~\cite{balle2017end}; it does not constrain the encoder's
sensitivity to input variations. The input-domain perturbation we
study acts upstream of $g_a$ and, as shown in
Section~\ref{sec:theory}, induces a structured penalty on the
encoder's Jacobian under the RD loss. The two mechanisms operate at
different stages and are complementary, and the input-side
perturbation is conceptually related to randomized
smoothing~\cite{cohen2019certified}.
A small concurrent line of work trains LIC with noisy inputs for the
explicit task of joint denoising and
compression~\cite{brummer2023importance}; the goal there is to
\emph{deploy} the codec on corrupted images, whereas we use noise
purely as a training-time regularizer and evaluate strictly on clean
inputs.

\section{Why Pixel-Domain Noise Helps: A Local RD Expansion}
\label{sec:theory}

We expand the noisy RD loss to second order in the noise scale
$\sigma$ and show that pixel-domain input noise turns the clean RD
objective into a
\emph{structured} regularizer with two interpretable components: an
information-weighted encoder Jacobian penalty (rate side) and a
Frobenius contractive penalty on the end-to-end reconstruction map
(distortion side). All algebra is collected in
Appendix~\ref{app:distortion}--\ref{app:summary}.

\subsection{Setup}
\label{sec:setup}

Let $g_a$, $g_s$ be the analysis and synthesis transforms. At
inference, $\mathbf{y}=g_a(\mathbf{x})$,
$\hat{\mathbf{y}}=Q(\mathbf{y})$,
$\hat{\mathbf{x}}=g_s(\hat{\mathbf{y}})$. During training we keep the
standard additive-uniform proxy $\mathbf{u}\!\sim\!\mathcal{U}(-\tfrac{1}{2},\tfrac{1}{2})$
for the latent and inject Gaussian noise only at the encoder input,
\begin{equation}
  \tilde{\mathbf{x}}=\mathbf{x}+\boldsymbol{\eta},
  \qquad
  \boldsymbol{\eta}\sim\mathcal{N}(0,\sigma^{2}I),
  \qquad
  \bar{\mathbf{y}}=g_a(\tilde{\mathbf{x}})+\mathbf{u}.
\end{equation}
The hyperprior variable is suppressed in the notation for clarity; the
additional $z$-pathway and context-model terms are made explicit in
Appendix~\ref{app:rate}. The noise-regularized training objective is then
\begin{equation}
  \mathcal{L}_{\mathrm{noise}}
  \;=\;
  \mathbb{E}_{\mathbf{x},\boldsymbol{\eta},\mathbf{u}}\!
  \bigl[\,R(\tilde{\mathbf{x}}) + \lambda\,D(\tilde{\mathbf{x}},\mathbf{x})\,\bigr],
  \;\;
  D=\|g_s(\bar{\mathbf{y}})-\mathbf{x}\|_2^{2},
  \;\;
  R=-\log p_\phi(\bar{\mathbf{y}}).
  \label{eq:noise_loss}
\end{equation}
With a slight abuse of notation, $R(\tilde{\mathbf{x}})$ should be
read as the composite end-to-end rate
$R(\tilde{\mathbf{x}}) := -\log p_\phi\!\bigl(g_a(\tilde{\mathbf{x}})+\mathbf{u}\bigr)$,
i.e.\ the cost of coding the noisy image's latent under the entropy
model. The key structural observation is that, because
$\boldsymbol{\eta}$ is injected \emph{upstream} of $g_a$, it
propagates through both the distortion path $g_s\!\circ\!g_a$ and
the rate path $-\log p_\phi(g_a(\cdot)+\mathbf{u})$. Compared with classical
denoising-autoencoder analyses~\cite{bishop1995training,rifai2011contractive},
where input noise perturbs only a single reconstruction loss, the
noisy LIC objective therefore receives \emph{two} second-order
corrections from $\boldsymbol{\eta}$, with different physical content.
The latent proxy $\mathbf{u}$ is held fixed throughout the
$\boldsymbol{\eta}$-expansion (it is independent of
$\boldsymbol{\eta}$ and enters only through
$\bar{\mathbf{y}}_0=g_a(\mathbf{x})+\mathbf{u}$;
see Appendix~\ref{app:rate}), so it generates no genuine cross terms.
Standard mild smoothness assumptions on $g_a, g_s, p_\phi$ suffice to
make the second-order Taylor remainder $O(\sigma^{4})$ under Gaussian
averaging; the precise statement, and a brief discussion of how
modern LIC backbones satisfy it in practice, is deferred to
Appendix~\ref{app:smoothness}.

\subsection{Distortion branch: end-to-end contraction with an LIC-specific burden}
\label{sec:dist_branch}

Define the training-time reconstruction map for a fixed quantization
proxy $\mathbf{u}$,
$f(\mathbf{x}) := g_s\bigl(g_a(\mathbf{x})+\mathbf{u}\bigr)$, and let
$D_0(\mathbf{x})=\|f(\mathbf{x})-\mathbf{x}\|_2^{2}$ be the clean-input
distortion. A second-order Taylor expansion in $\boldsymbol{\eta}$
followed by Gaussian averaging gives (Appendix~\ref{app:distortion})
\begin{equation}
  \mathbb{E}_{\boldsymbol{\eta}}\!\bigl[D(\tilde{\mathbf{x}},\mathbf{x})\bigr]
  \;=\;
  D_0(\mathbf{x})
  \;+\;
  \sigma^{2}\,\|J_f(\mathbf{x})\|_F^{2}
  \;+\;
  \sigma^{2}\,\bigl\langle f(\mathbf{x})-\mathbf{x},\,\Delta f(\mathbf{x})\bigr\rangle
  \;+\; O(\sigma^{4}),
  \label{eq:distortion_expansion_main}
\end{equation}
where $\Delta f$ is the component-wise Laplacian of $f$. The first
correction $\sigma^{2}\|J_f\|_F^{2}$ is the LIC analogue of the
contractive autoencoder penalty~\cite{rifai2011contractive} and
upper-bounds the squared spectral norm $\|J_f\|_2^{2}$, serving as the
natural surrogate of the codec's local Lipschitz constant; we measure
the contractions of $J_{g_a}$ and $J_f$ directly in
Section~\ref{sec:jacobian}. The second correction is a residual--Hessian
interaction \emph{specific to lossy compression}: a classical DAE
satisfies $f^\star\!\approx\!\mathbf{x}$ at the optimum and the term
collapses to $O(\sigma^{4})$, but the rate constraint in LIC keeps
$f-\mathbf{x}\!\neq\!0$ at any finite bitrate, so the term survives
at $O(\sigma^{2})$, is sign-indefinite, and is most pronounced at low
rate.

\subsection{Rate branch: an information-weighted encoder Jacobian}
\label{sec:rate_branch}

Let $\bar{\mathbf{y}}_0:=g_a(\mathbf{x})+\mathbf{u}$ denote the
clean-input training-time latent, and let
$s_\phi := \nabla\log p_\phi$ and
$I_\phi := -\nabla^{2}\log p_\phi$ be the score and observed
information of the entropy model. Expanding the rate
$R(\tilde{\mathbf{x}}) = -\log p_\phi(\bar{\mathbf{y}})$ to second
order around $\bar{\mathbf{y}}_0$ in the latent perturbation
$\delta\mathbf{y} = J_{g_a}(\mathbf{x})\boldsymbol{\eta}
   + \tfrac{1}{2}H_{g_a}(\mathbf{x})[\boldsymbol{\eta},\boldsymbol{\eta}]
   + O(\|\boldsymbol{\eta}\|^{3})$
and averaging over $\boldsymbol{\eta}$ yields
(Appendix~\ref{app:rate})
\begin{equation}
  \mathbb{E}_{\boldsymbol{\eta}}\!\bigl[R(\tilde{\mathbf{x}})\bigr]
  \;=\;
  R_0(\mathbf{x})
  \;+\;
  \tfrac{\sigma^{2}}{2}\,
  \mathrm{tr}\!\bigl(J_{g_a}(\mathbf{x})^{\!\top}\,I_\phi(\bar{\mathbf{y}}_0)\,J_{g_a}(\mathbf{x})\bigr)
  \;-\;
  \tfrac{\sigma^{2}}{2}\,s_\phi(\bar{\mathbf{y}}_0)^{\!\top}\,\Delta g_a(\mathbf{x})
  \;+\; O(\sigma^{4}),
  \label{eq:rate_expansion_main}
\end{equation}
with $R_0(\mathbf{x}) := -\log p_\phi(\bar{\mathbf{y}}_0)$ denoting the
clean-input rate and $\Delta g_a$ denoting the component-wise Laplacian of $g_a$. The Mahalanobis
form $\mathrm{tr}(J_{g_a}^{\!\top} I_\phi J_{g_a})$ is an
\emph{information-weighted} encoder contraction whose strength tracks
the rate landscape automatically: for standard LIC entropy models,
$I_\phi$ is positive semi-definite, so this term acts as a genuine
penalty; where $p_\phi$ is sharp, $I_\phi$ is
large and the encoder is strongly penalized for being sensitive to
input variation \emph{in those very directions}, while in flatter
regions the effect correspondingly weakens. This is the LIC-native
counterpart of the contractive penalty of~\cite{rifai2011contractive},
but with no extra hyperparameter beyond $\sigma$. The second correction
$-\,s_\phi(\bar{\mathbf{y}}_0)^{\!\top}\Delta g_a$ is a rate-side
score--Laplacian term, locally sign-indefinite; at a well-calibrated
entropy model it oscillates around zero and is dominated by the
sign-definite $\mathrm{tr}(J_{g_a}^{\!\top} I_\phi J_{g_a})$, so we
focus the interpretation on the latter while keeping it explicit in
Eq.~\eqref{eq:combined_regularizer}.

\subsection{Combined regularizer}
\label{sec:combined}

Substituting Eqs.~\eqref{eq:distortion_expansion_main}--\eqref{eq:rate_expansion_main}
into Eq.~\eqref{eq:noise_loss} yields the local form of the noisy
RD loss as a clean-input RD term plus four explicit $O(\sigma^{2})$
corrections,
\begin{equation}
  \mathcal{L}_{\mathrm{noise}}(\mathbf{x})
  \;=\;
  \mathcal{L}_0(\mathbf{x})
  \;+\;
  \tfrac{\sigma^{2}}{2}\!\left(\Omega_R^{(1)} + \Omega_R^{(2)}\right)
  \;+\;
  \lambda\sigma^{2}\!\left(\Omega_D^{(1)} + \Omega_D^{(2)}\right)
  \;+\; O(\sigma^{4}),
  \label{eq:combined_regularizer}
\end{equation}
\begin{equation}
  \begin{aligned}
    \Omega_R^{(1)}
    &= \mathrm{tr}\!\bigl(J_{g_a}^{\!\top} I_\phi(\bar{\mathbf{y}}_0) J_{g_a}\bigr),
    & \quad
    \Omega_R^{(2)}
    &= -\,s_\phi(\bar{\mathbf{y}}_0)^{\!\top}\Delta g_a(\mathbf{x}), \\[2pt]
    \Omega_D^{(1)}
    &= \|J_f(\mathbf{x})\|_F^{2},
    & \quad
    \Omega_D^{(2)}
    &= \bigl\langle f(\mathbf{x})-\mathbf{x},\,\Delta f(\mathbf{x})\bigr\rangle.
  \end{aligned}
  \label{eq:omega_definitions}
\end{equation}
The headline regularizers are the two sign-definite penalties
$\Omega_R^{(1)}$ and $\Omega_D^{(1)}$, which actively shape the learned
codec and are measured directly in Section~\ref{sec:jacobian}; the
sign-indefinite $\Omega_R^{(2)},\Omega_D^{(2)}$ are kept explicit for
completeness, with $\Omega_D^{(2)}$ acting as the LIC-specific
residual--Hessian burden that the rate constraint cannot make vanish
at finite bitrate.

\paragraph{Spectral and manifold interpretation.}
$\Omega_R^{(1)}$ rescales each input direction by the entropy-model
curvature along its image, $\langle J_{g_a}\mathbf{v}_i,\,I_\phi
J_{g_a}\mathbf{v}_i\rangle$, where $\mathbf{v}_i$ are the right
singular vectors of $J_{g_a}$. For the mean-scale Gaussian
conditionals used in standard LIC entropy models, $I_\phi$ is
diagonal with entries equal to the inverse predicted variances, which
the hyperprior adapts spatially: directions the model commits to
encoding precisely carry large $I_\phi$ and are penalized most. The
result is an \emph{anisotropic} encoder that retains on-manifold
variation and suppresses off-manifold sensitivity, formalizing the
manifold-shaping intuition behind the recipe; we test this
interpretation in Section~\ref{sec:jacobian}.

\paragraph{Connections to information bottleneck and score matching.}
Two complementary perspectives further support this picture. An
information-bottleneck~\cite{tishby2000information} argument shows
that minimizing $\mathcal{L}_{\mathrm{noise}}$ pushes the encoder
toward conditional independence of $\mathbf{y}$ from
$\boldsymbol{\eta}$ given $\mathbf{x}$, mechanistically consistent
with $\Omega_R^{(1)}$ shrinking the very directions through which
$\boldsymbol{\eta}$ leaks into the latent. A score matching
argument shows that, absent the rate term, the optimum reduces to
a classical DAE whose residual aligns at $O(\sigma^{2})$ with the
Stein score of the noise-smoothed data
distribution~\cite{vincent2011connection,alain2014regularized}; the
rate constraint perturbs this stationary point at $O(\sigma^{2})$
through the LIC-specific scalar penalty $\Omega_D^{(2)}$. The full
derivations are deferred to Appendix~\ref{app:connections}.

\section{Method}
\label{sec:method}

\paragraph{Standard pipeline.}
Given $\mathbf{x}\!\in\!\mathbb{R}^{H\times W\times 3}$, the encoder
produces $\mathbf{y}=g_a(\mathbf{x})$, which is rounded to
$\hat{\mathbf{y}}=\lfloor\mathbf{y}\rceil$ at inference and replaced by
$\tilde{\mathbf{y}}=\mathbf{y}+\mathcal{U}(-\tfrac{1}{2},\tfrac{1}{2})$
during training~\cite{balle2017end}. A hyperprior
$\mathbf{z}=h_a(\mathbf{y})$ parameterizes the entropy model
$p_{\hat{\mathbf{y}}|\hat{\mathbf{z}}}$, and the decoder returns
$\hat{\mathbf{x}}=g_s(\hat{\mathbf{y}})$. The standard objective is
\(
  \mathcal{L}_{\mathrm{std}} = R + \lambda\,d(\mathbf{x},\hat{\mathbf{x}})
\)
with $R = \mathbb{E}[-\log_2 p_{\hat{\mathbf{y}}|\hat{\mathbf{z}}} - \log_2 p_{\hat{\mathbf{z}}}]$.

\paragraph{Noise-regularized objective.}
Following Section~\ref{sec:theory}, we replace the clean encoder input
by a clamped Gaussian-perturbed version while keeping the distortion
target clean:
\begin{equation}
  \boldsymbol{\eta}\!\sim\!\mathcal{N}(0,\sigma^{2}I),\quad
  \tilde{\mathbf{x}} = \mathrm{clamp}(\mathbf{x}+\boldsymbol{\eta},\,0,\,1),\quad
  \mathcal{L}_{\mathrm{noise}}
  = R(\tilde{\mathbf{x}}) + \lambda\,d\!\bigl(\mathbf{x},\,g_s(\bar{\mathbf{y}})\bigr),
  \label{eq:loss}
\end{equation}
with $\bar{\mathbf{y}} = g_a(\tilde{\mathbf{x}}) + \mathbf{u}$ as in
Section~\ref{sec:setup}. The noisy image is fed into \emph{both} the
analysis transform and the hyperprior pathway. The local expansions are
understood on the unclamped region. Clamping keeps training inputs in
range, and its boundary effect diminishes as $\sigma$ is annealed; a
no-clamp ablation gives similar RD performance
(Appendix~\ref{app:clamping}).

\paragraph{Noise schedule.}
Annealing $\sigma$ from a large $\sigma_0$ down to $0$ separates the
geometric-shaping role of $(\Omega_R^{(1)},\Omega_D^{(1)})$ from the
persistent low-rate cost of $\Omega_D^{(2)}$: a large $\sigma$ early
in training biases the encoder--decoder along the data manifold, and
$\sigma\!\to\!0$ later sheds all four $O(\sigma^{2})$ corrections at
once, returning optimization to the clean RD loss
(Appendix~\ref{app:annealing}). Concretely, with $T$ noise-active
epochs and initial scale $\sigma_0$, we use the exponential schedule
\begin{equation}
  \sigma(t) \;=\; \sigma_0\,\exp\!\bigl(-k\,t/(T-1)\bigr),
  \quad k = \ln 100 \approx 4.6,
  \label{eq:schedule}
\end{equation}
so $\sigma(T\!-\!1)\!\approx\!0.01\,\sigma_0$ and $\sigma(t)=0$ for
$t\ge T$. Within each epoch, individual minibatches draw a noise
level $\sigma_{\mathrm{batch}} = r\cdot\sigma(t)$ with
$r\!\sim\!\mathcal{U}(0,1)$ (a fresh scalar per minibatch, distinct
from the latent proxy $\mathbf{u}$), so the model sees a continuum
of noise scales at every point of the schedule rather than a single
instantaneous value.

\paragraph{Algorithm.}
Algorithm~\ref{alg:nat} summarizes the recipe; the only change from
standard LIC training is to perturb the encoder input and keep the
reconstruction target clean. A minimal PyTorch sketch with a drop-in
\texttt{lic\_noise} helper is given in Appendix~\ref{app:lic_noise}.

\begin{algorithm}[t]
\caption{Noise-Regularized Training for LIC}
\label{alg:nat}
\begin{algorithmic}[1]
\REQUIRE Codec $(g_a,g_s,h_a,h_s)$, dataset $\mathcal{X}$,
         initial noise $\sigma_0$, noise epochs $T$, total epochs $T_{\mathrm{total}}$,
         RD trade-off $\lambda$ (a constant for single-rate codecs;
         drawn from a logarithmic grid $\lambda\!\in\![16,4096]$ over
         $64$ QPs per minibatch for variable-rate codecs)
\FOR{epoch $t=0,\dots,T_{\mathrm{total}}-1$}
  \STATE Compute $\sigma(t)$ via Eq.~\eqref{eq:schedule}; set $\sigma(t)\!=\!0$ if $t\ge T$
  \FOR{each minibatch $\mathbf{x}\in\mathcal{X}$}
    \STATE Sample $r\!\sim\!\mathcal{U}(0,1)$;\; $\sigma_{\mathrm{batch}}=r\cdot\sigma(t)$ \hfill $\triangleright$ schedule multiplier (distinct from latent $\mathbf{u}$)
    \IF{$\sigma_{\mathrm{batch}} > 0$}
      \STATE $\boldsymbol{\eta}\!\sim\!\mathcal{N}(0,\sigma_{\mathrm{batch}}^2 I)$;\;
             $\tilde{\mathbf{x}}\leftarrow\mathrm{clamp}(\mathbf{x}+\boldsymbol{\eta},0,1)$
    \ELSE
      \STATE $\tilde{\mathbf{x}}\leftarrow\mathbf{x}$
    \ENDIF
    \STATE $\bar{\mathbf{y}}\leftarrow g_a(\tilde{\mathbf{x}}) + \mathbf{u}$;\;
           $\hat{\mathbf{x}}\leftarrow g_s(\bar{\mathbf{y}})$;\;
           $\mathcal{L}\leftarrow R(\tilde{\mathbf{x}}) + \lambda\,d(\mathbf{x},\hat{\mathbf{x}})$
           \hfill $\triangleright$ clean target
    \STATE Update $\theta$ with Adam on $\nabla_\theta\mathcal{L}$
  \ENDFOR
\ENDFOR
\end{algorithmic}
\end{algorithm}

\section{Experiments}
\label{sec:experiments}

\subsection{Experimental Setup}

\paragraph{Backbones.}
We evaluate the recipe on two representative codecs that span both
operating regimes used in the field. For the \emph{single-rate}
setting we use ELIC~\cite{he2022elic}; for the \emph{variable-rate}
setting we use the intra image codec of
DCVC-RT~\cite{jia2025dcvcrt}, referred to as DMCI in the literature.
We reproduce DMCI at three model scales, denoted
\emph{DMCI-Base}, \emph{DMCI-Medium}, and \emph{DMCI-Large}, obtained
by jointly scaling three channel widths: the analysis/synthesis
transform $\mathrm{ch}_g$, the latent $\mathbf{y}$ dimension
$\mathrm{ch}_y$, and the hyperprior $\mathbf{z}$ dimension
$\mathrm{ch}_z$. Concretely,
$(\mathrm{ch}_g,\,\mathrm{ch}_y,\,\mathrm{ch}_z)
=(368,\,256,\,128)$ for Base,
$(512,\,368,\,192)$ for Medium, and
$(736,\,512,\,256)$ for Large; all other architectural blocks remain identical across scales.

\paragraph{Training.}
All models are trained on MLIC-Train-100K~\cite{mlictrain100k} with
Adam~\cite{kingma2015adam}. For \emph{single-rate} ELIC, we use an
initial learning rate of $1\!\times\!10^{-4}$, decayed by a factor of
$10$ at $\sim\!85\%$ and $\sim\!99\%$ of the total $\sim\!2$M
iterations. The first $90\%$ uses batch size $32$ and patch size $256$;
the remaining $10\%$ uses batch size $16$ and patch size $512$. For
\emph{variable-rate} DMCI, we start from a fully trained single-rate
counterpart and fine-tune in two stages: $1.4$M iterations with the
noise-regularized objective \eqref{eq:loss} at learning rate
$1\!\times\!10^{-4}$, followed by $0.1$M iterations at patch size $512$
and learning rate $1\!\times\!10^{-5}$. Each stage applies a single
factor-of-ten decay late in training. Throughout DMCI training, the RD
trade-off $\lambda$ is sampled per minibatch from a logarithmic grid of
$64$ QPs covering $[16,4096]$, and the distortion target remains the
clean image.

\paragraph{Datasets and metrics.}
We evaluate on Kodak~\cite{kodak} ($24$ images, $512\!\times\!768$),
CLIC Professional Validation~\cite{clic} ($41$ images), and
Tecnick~\cite{tecnick} ($100$ images, $1200\!\times\!1200$). All
results are reported as
BD-Rate~\cite{bjontegaard2001calculation} savings (negative means
bitrate savings at matched reconstruction quality). All baseline and
noise-regularized numbers are reproduced under the same training
pipeline unless otherwise noted.

\subsection{RD Performance}

\paragraph{Within-backbone gain (Table~\ref{tab:bdrate_results}).}
On the single-rate ELIC backbone, noise-regularized training reduces
BD-Rate by $2.10\%$ (Kodak), $3.28\%$ (Tecnick), and $2.42\%$ (CLIC),
with a $2.60\%$ average gain over the reproduced clean-trained
baseline. On the variable-rate DMCI backbone, the same recipe yields
consistent gains across all three scales: $-0.85\%$, $-1.05\%$, and
$-2.38\%$ for Base, Medium, and Large, respectively. The largest gain
appears on DMCI-Large, suggesting that the regularizer becomes more
beneficial as model capacity increases. The official open-source
release of DMCI-Base is reported here only for reference; it is
$+4.66\%$ worse than our reproduced baseline under the same evaluation
setup. Appendix~\ref{app:hpcm} additionally reports a preliminary
measurement on HPCM-Large~\cite{li2025hpcm}, where the noise-regularized
variant lowers bitrate by $3.7$--$4.6\%$ across Kodak, CLIC, and Tecnick
at a slight PSNR degradation; a full BD-Rate sweep on this backbone is deferred
to follow-up work.

\begin{table}[t]
  \caption{BD-Rate on Kodak, Tecnick, and CLIC Pro Valid.
           Each sub-group is anchored to its own baseline ($0.00\%$).
           All models reproduced under a unified training pipeline
           unless otherwise noted. Negative: bitrate savings at matched PSNR.}
  \label{tab:bdrate_results}
  \centering
  \small
  \setlength{\tabcolsep}{2.6pt}
  \begin{tabular*}{0.9\columnwidth}{@{\extracolsep{\fill}}p{0.18\columnwidth}lrrrr@{}}
    \toprule
    Model & Variant & Kodak & Tecnick & CLIC & Avg. \\
    \midrule
    \multirow{2}{*}{ELIC}
      & baseline                          & $0.00\%$ & $0.00\%$ & $0.00\%$ & $0.00\%$ \\
      & \hspace{0.5em}w/\,Noise\,Reg.\ (Ours) & $\mathbf{-2.10\%}$ & $\mathbf{-3.28\%}$ & $\mathbf{-2.42\%}$ & $\mathbf{-2.60\%}$ \\
    \midrule
    \multirow{7}{*}{DMCI}
      & Base (baseline)                   & $0.00\%$ & $0.00\%$ & $0.00\%$ & $0.00\%$ \\
      & \hspace{0.5em}w/\,Noise\,Reg.\ (Ours) & $\mathbf{-0.90\%}$ & $\mathbf{-0.85\%}$ & $\mathbf{-0.80\%}$ & $\mathbf{-0.85\%}$ \\
      & \hspace{0.5em}official release$^\dagger$ & $+2.91\%$ & $+5.60\%$ & $+5.45\%$ & $+4.66\%$ \\
    \cmidrule(lr){2-6}
      & Medium (baseline)                 & $0.00\%$ & $0.00\%$ & $0.00\%$ & $0.00\%$ \\
      & \hspace{0.5em}w/\,Noise\,Reg.\ (Ours) & $\mathbf{-0.98\%}$ & $\mathbf{-0.93\%}$ & $\mathbf{-1.24\%}$ & $\mathbf{-1.05\%}$ \\
    \cmidrule(lr){2-6}
      & Large (baseline)                  & $0.00\%$ & $0.00\%$ & $0.00\%$ & $0.00\%$ \\
      & \hspace{0.5em}w/\,Noise\,Reg.\ (Ours) & $\mathbf{-2.32\%}$ & $\mathbf{-2.32\%}$ & $\mathbf{-2.50\%}$ & $\mathbf{-2.38\%}$ \\
    \bottomrule
  \end{tabular*}
  \vspace{3pt}
  \parbox{\linewidth}{\footnotesize
    $^\dagger$~Official open-source release of DMCI-Base;
    not reproduced under our unified training pipeline.}
\end{table}

Across eight DMCI-Base runs with $\sigma_0\in[0.4,0.8]$, every run
improves over the shared clean baseline on all three datasets, with
within-$\sigma_0$ standard deviation at most $0.04$ percentage
points.

\paragraph{Sensitivity to the initial noise scale (Table~\ref{tab:sigma_ablation}).}
The recipe exposes a single hyperparameter, the initial noise scale
$\sigma_0$ in Eq.~\eqref{eq:schedule}. Sweeping $\sigma_0$ on the
DMCI-Base backbone (Kodak) shows that very small values
(e.g.\ $0.125$) collapse to the no-noise solution at $+0.01\%$, while
the BD-Rate saturates around $\sigma_0\!\in\![0.5,\,0.75]$ at roughly
$-0.9\%$, with $\sigma_0\!=\!0.5$ giving the best result. The narrow
spread across the entire sweep ($\sim\!0.9$\,pp from worst to best)
indicates that sufficiently large initial scales yield similar benefits;
we therefore recommend $\sigma_0\!\in\![0.4,0.8]$.

\begin{table}[t]
  \caption{Sensitivity of noise-regularized DMCI-Base to the initial
           noise scale $\sigma_0$ on Kodak. BD-Rate anchored to
           our reproduced DMCI-Base (clean training).}
  \label{tab:sigma_ablation}
  \centering
  \small
  \setlength{\tabcolsep}{10pt}
  \begin{tabular}{lcccc}
    \toprule
    Initial noise scale $\sigma_0$ & $0.125$ & $0.25$ & $0.50$ & $0.75$ \\
    \midrule
    BD-Rate on Kodak & $+0.01\%$ & $-0.16\%$ & $\mathbf{-0.90\%}$ & $-0.86\%$ \\
    \bottomrule
  \end{tabular}
\end{table}

\paragraph{Comparison against VTM-22.0 and prior codecs (Table~\ref{tab:vtm_main}).}
Table~\ref{tab:vtm_main} situates our trained models alongside VTM-22.0
and a broader set of recent LIC systems---TCM~\cite{liu2023tcm},
MLIC++~\cite{mlicpp}, LALIC~\cite{feng2025lalic}, DCAE~\cite{dcae},
and HPCM~\cite{li2025hpcm}---reported as absolute BD-Rate against
VTM-22.0. Noise-regularized ELIC improves the reproduced ELIC baseline
from $-6.13\%$ to $-8.71\%$ average BD-Rate, a further $2.58$\,pp gain
over VTM-22.0 even on this relatively lightweight baseline. The three
DMCI scales reach $-18.11\%$, $-20.13\%$, and $-22.02\%$ with the
proposed training; in particular, DMCI-Medium w/ Noise Reg.\ already
surpasses HPCM-Large ($-19.86\%$) while using $953$ vs.\ $1261$
kMACs/pixel. 
While this absolute comparison serves as a benchmark snapshot, the isolated marginal gain from noise-regularized training is reported in the within-backbone column of Table~\ref{tab:bdrate_results} (RD curves in Appendix~\ref{app:rd_curves}).

\begin{table*}[t]
  \caption{Main RD comparison against VTM-22.0 in BD-Rate on Kodak,
           Tecnick, and CLIC. Negative values indicate bitrate savings
           at matched PSNR. Our noise-regularized variants are
           highlighted in bold; the best overall result is additionally
           bolded in the metric columns.}
  \label{tab:vtm_main}
  \centering
  \small
  \setlength{\tabcolsep}{4pt}
  \resizebox{\textwidth}{!}{%
  \begin{tabular}{lcccrrrr}
    \toprule
    Method & kMACs/pixel & Param. (M) & Type & Kodak & Tecnick & CLIC & Avg. \\
    \midrule
    TCM & 1824 & 77 & Single-Rate & $-10.46\%$ & $-10.91\%$ & $-9.23\%$ & $-10.20\%$ \\
    MLIC++ & 1283 & 117 & Single-Rate & $-13.61\%$ & $-17.18\%$ & $-14.50\%$ & $-15.10\%$ \\
    LALIC & 667 & 66 & Single-Rate & $-14.01\%$ & $-17.71\%$ & $-15.61\%$ & $-15.78\%$ \\
    DCAE & 940 & 119 & Single-Rate & $-15.72\%$ & $-20.16\%$ & $-17.13\%$ & $-17.67\%$ \\
    HPCM-Base & 919 & 69 & Single-Rate & $-14.67\%$ & $-17.37\%$ & $-15.26\%$ & $-15.77\%$ \\
    HPCM-Large & 1261 & 90 & Single-Rate & $-18.70\%$ & $-21.44\%$ & $-19.43\%$ & $-19.86\%$ \\
    \midrule
    ELIC (reproduced) & 578 & 32 & Single-Rate & $-5.87\%$ & $-8.11\%$ & $-4.42\%$ & $-6.13\%$ \\
    \textbf{ELIC w/ Noise Reg.} & 578 & 32 & Single-Rate & $\mathbf{-7.98\%}$ & $\mathbf{-11.35\%}$ & $\mathbf{-6.79\%}$ & $\mathbf{-8.71\%}$ \\
    DMCI-Base (official) & 486 & 46 & Variable-Rate & $-12.67\%$ & $-15.38\%$ & $-12.63\%$ & $-13.56\%$ \\
    DMCI-Base & 486 & 46 & Variable-Rate & $-15.23\%$ & $-19.87\%$ & $-16.99\%$ & $-17.36\%$ \\
    \textbf{DMCI-Base w/ Noise Reg.} & 486 & 46 & Variable-Rate & $\mathbf{-16.04\%}$ & $\mathbf{-20.59\%}$ & $\mathbf{-17.70\%}$ & $\mathbf{-18.11\%}$ \\
    DMCI-Medium & 953 & 91 & Variable-Rate & $-17.37\%$ & $-21.85\%$ & $-18.45\%$ & $-19.23\%$ \\
    \textbf{DMCI-Medium w/ Noise Reg.} & 953 & 91 & Variable-Rate & $\mathbf{-18.24\%}$ & $\mathbf{-22.63\%}$ & $\mathbf{-19.53\%}$ & $\mathbf{-20.13\%}$ \\
    DMCI-Large & 1924 & 181 & Variable-Rate & $-18.50\%$ & $-22.67\%$ & $-18.93\%$ & $-20.03\%$ \\
    \textbf{DMCI-Large w/ Noise Reg.} & 1924 & 181 & Variable-Rate & $\mathbf{-20.47\%}$ & $\mathbf{-24.55\%}$ & $\mathbf{-21.06\%}$ & $\mathbf{-22.02\%}$ \\
    \bottomrule
  \end{tabular}%
  }
\end{table*}

\begin{figure}[t]
  \centering
  \includegraphics[width=\linewidth]{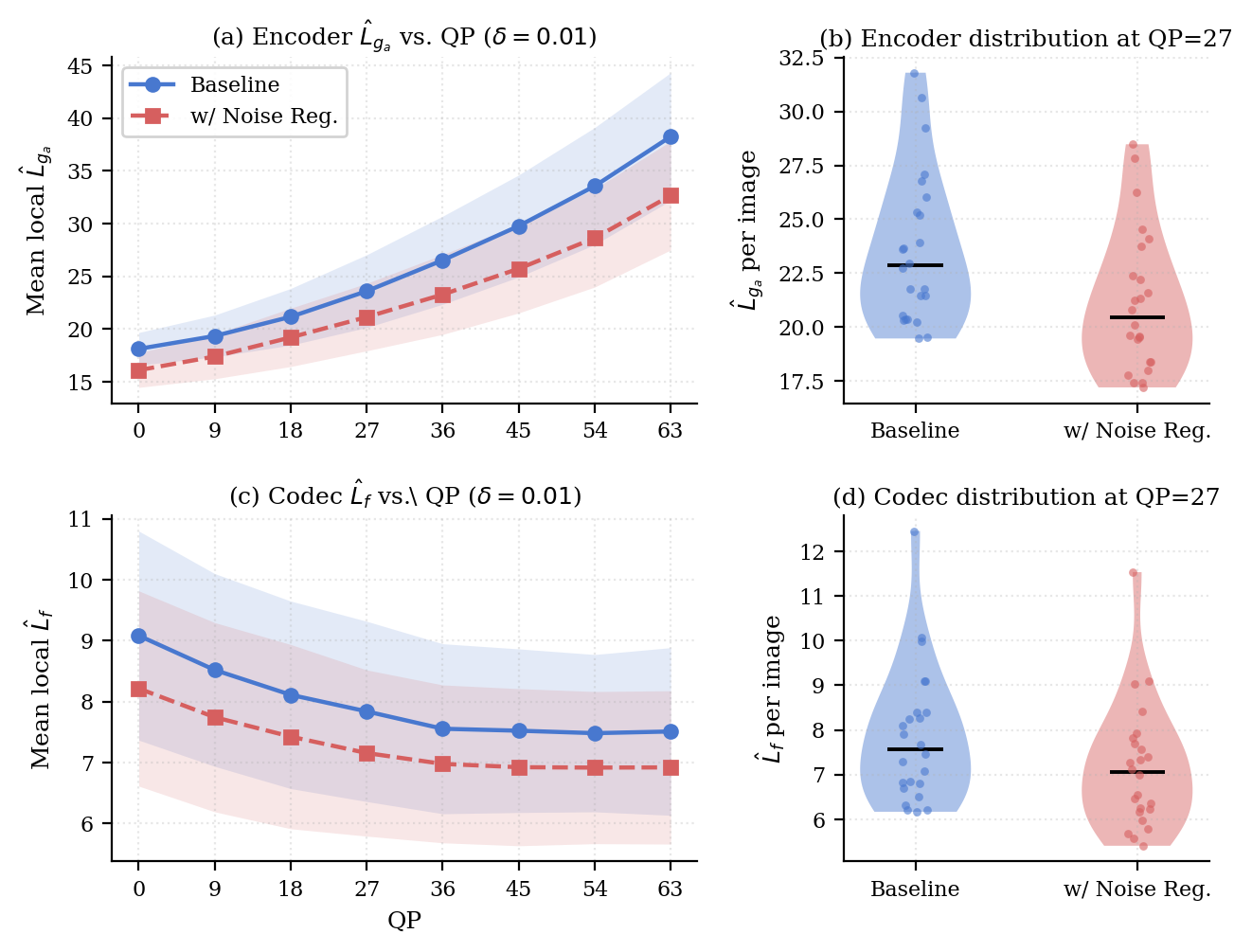}
  \caption{%
    \textbf{Local Lipschitz analysis on Kodak (DMCI-Base,
    $\delta\!=\!0.01$).}
    Top row, encoder $\hat L_{g_a}$ (tests $\Omega_R^{(1)}$);
    bottom row, end-to-end codec $\hat L_f$ on the
    quantization-free map $f=g_s\!\circ\!g_a$ (tests
    $\Omega_D^{(1)}=\|J_f\|_F^{2}$).
    Left panels~(a,c): mean $\hat{L}$ vs.\ QP for the baseline and
    the noise-regularized model. Right panels~(b,d): per-image distribution at
    QP$\,=\,27$, estimated with $n\!=\!64$ random directions.
    Noise regularization contracts both Jacobians at every operating
    point ($12.0\%$ encoder, $8.4\%$ codec on average; see
    Table~\ref{tab:lipschitz_summary}).
  }
  \label{fig:lipschitz}
\end{figure}

\subsection{Jacobian Contraction and Information Weighting}
\label{sec:jacobian}

The local objective \eqref{eq:combined_regularizer} predicts both
contraction and its directional structure: the rate-side penalty
$\Omega_R^{(1)}=\mathrm{tr}(J_{g_a}^{\!\top}I_\phi J_{g_a})$ contracts
the \emph{encoder} Jacobian preferentially along high-$I_\phi$
coordinates, while the distortion-side penalty
$\Omega_D^{(1)}=\|J_f\|_F^{2}$ contracts the \emph{end-to-end}
reconstruction map $f=g_s\!\circ\!g_a$. We first test both contractions
on the variable-rate DMCI-Base backbone with the same randomized
finite-difference protocol applied to two probe functions
$\Phi\in\{g_a,\,f\}$. For $n\!=\!64$ random unit-norm directions
$\mathbf{d}_i$ in pixel space and perturbation magnitude $\delta$, we
define a finite-direction local sensitivity estimate
\begin{equation}
  \hat{L}_{\Phi}(\mathbf{x},\mathrm{QP})
  \;=\;
  \max_i\;
  \frac{\bigl\|\Phi(\mathbf{x}+\delta\mathbf{d}_i;\mathbf{q}_{\mathrm{enc}}) - \Phi(\mathbf{x};\mathbf{q}_{\mathrm{enc}})\bigr\|_{\mathrm{rms}}}
       {\delta\,\|\mathbf{d}_i\|_{\mathrm{rms}}},
  \qquad
  \Phi\in\{g_a,\,f\},
  \label{eq:lip_estimate}
\end{equation}
where $\mathbf{q}_{\mathrm{enc}}$ is the QP-conditioning vector of the
variable-rate codec and
$\|\cdot\|_{\mathrm{rms}} = \|\cdot\|_F/\sqrt{\mathrm{numel}}$ is the
RMS norm. We use $f$ in its differentiable, quantization-free form
$f(\mathbf{x})=g_s\bigl(g_a(\mathbf{x})\bigr)$ ($\mathbf{u}\!=\!0$),
matching the theoretical map of Section~\ref{sec:dist_branch}. RMS
normalization removes the output-dimensionality factor and facilitates
comparison across QPs. We sweep
$\delta\in\{0.005,0.01,0.02,0.05\}$ and treat as robust only reductions
that persist across the entire grid.

\paragraph{Encoder contraction (verifying $\Omega_R^{(1)}$).}
Figure~\ref{fig:lipschitz}\,(a,b) reports $\hat{L}_{g_a}$ at the focal
probe scale $\delta\!=\!0.01$. The noise-regularized encoder is
\emph{consistently less sensitive} at every operating point, with an
average $12.0\%$ reduction across QPs (per-QP range
$9.2\%$--$14.7\%$). For concreteness, $\hat L_{g_a}$ contracts from
$18.12\!\to\!16.10$ at QP\,0 and from $38.25\!\to\!32.66$ at QP\,63.
Both models follow the expected monotone increase of $\hat{L}_{g_a}$
with QP, since higher rates demand finer spatial discrimination; noise
regularization reduces this sensitivity across all operating points.

\paragraph{Which encoder directions contract?}
The $12\%$ reduction is an overall average: it does not tell us whether
all directions shrink equally or some shrink more. To examine this
directional structure at a larger model scale, we next evaluate exact
JVPs on DMCI-Large and compare the same encoder response with and
without entropy-model weighting. For each direction, let
$v=J_{g_a}d$; $u=\|v\|^2$ measures its plain response energy, while
$p=v^\top I_\phi v$ weights that response by coding curvature. We sum
these energies over 24 Kodak images and the same $n=64$ directions per
image and QP, using a shared noise-model $I_\phi$, and define
\begin{equation}
  R_u=\frac{\sum_{\mathbf{x},d}u_{\mathrm{noise}}}
            {\sum_{\mathbf{x},d}u_{\mathrm{base}}},
  \qquad
  R_w=\frac{\sum_{\mathbf{x},d}p_{\mathrm{noise}}}
            {\sum_{\mathbf{x},d}p_{\mathrm{base}}}.
  \label{eq:info_ratios}
\end{equation}
A positive gap $R_u-R_w$ means that high-$I_\phi$ directions contract
more strongly. Table~\ref{tab:info_weighting} shows this pattern at
every QP.

\begin{table*}[t]
  \caption{Encoder information-weighting test on DMCI-Large with exact
           JVPs and a shared noise-model $I_\phi$. Weighted contraction
           exceeds unweighted contraction at every QP, yielding a
           positive gap of $7.70$--$17.72$ percentage points.}
  \label{tab:info_weighting}
  \centering
  \small
  \setlength{\tabcolsep}{3pt}
  \begin{tabular}{lrrrrrrrr}
    \toprule
    QP & 0 & 9 & 18 & 27 & 36 & 45 & 54 & 63 \\
    \midrule
    Weighted $1-R_w$ & $39.15\%$ & $41.41\%$ & $46.56\%$ & $50.99\%$ & $46.08\%$ & $40.52\%$ & $38.05\%$ & $38.03\%$ \\
    Unweighted $1-R_u$ & $29.75\%$ & $27.77\%$ & $31.13\%$ & $33.27\%$ & $32.28\%$ & $30.52\%$ & $30.13\%$ & $30.33\%$ \\
    Gap (pp) & $+9.41$ & $+13.64$ & $+15.43$ & $+17.72$ & $+13.80$ & $+10.00$ & $+7.91$ & $+7.70$ \\
    \bottomrule
  \end{tabular}
\end{table*}

\paragraph{End-to-end codec contraction (verifying $\Omega_D^{(1)}$).}
We next return to overall sensitivity, applying the same protocol to the
complete encoder--decoder map $\Phi=f$ to probe the distortion-side
penalty $\|J_f\|_F^{2}$ directly
(Figure~\ref{fig:lipschitz}\,(c,d)). Noise regularization again
contracts the codec at every operating point, with an $8.4\%$
average reduction (per-QP range $7.4$--$10.0\%$). Two qualitative
differences with the encoder result are worth noting. First,
$\hat L_f$ is both \emph{smaller} in absolute scale ($9.08$ at QP\,0
vs.\ $18.12$ for $\hat L_{g_a}$) and \emph{less QP-dependent}
($9.08\!\to\!7.51$ from QP\,0 to QP\,63), consistent with the
synthesis transform $g_s$ partially undoing the encoder's
amplification. Second, the relative reduction is \emph{stronger at
low rate} (QP\,0: $-9.6\%$) and weakens slowly at high rate (QP\,63:
$-7.9\%$), the opposite of the encoder trend.

\paragraph{Summary.}
Both contractions persist across an order-of-magnitude sweep of
$\delta$ (Table~\ref{tab:lipschitz_summary}; full per-QP, per-$\delta$
numbers in Appendix~\ref{app:lipschitz_full}). The JVP result adds a
simple directional picture: the encoder does not shrink all directions
equally, but contracts high-$I_\phi$ directions more strongly. Together,
these measurements capture both the magnitude and directional structure
of the local RD regularizer derived in Section~\ref{sec:theory}.

\begin{table}[t]
  \centering
  \small
  \caption{%
    \textbf{Summary of relative reduction in local Lipschitz
    $\hat L$ on Kodak DMCI-Base (baseline $\to$ noise-regularized).}
    Encoder column tests $\Omega_R^{(1)}$; end-to-end column
    tests $\Omega_D^{(1)}$. Each cell averages
    Eq.~\eqref{eq:lip_estimate} over $24$ Kodak images, $8$ QPs, and $n\!=\!64$ random directions. Full per-QP per-$\delta$ tables are in
    Appendix~\ref{app:lipschitz_full}.}
  \label{tab:lipschitz_summary}
  \setlength{\tabcolsep}{6pt}
  \begin{tabular}{l|cccc|c}
    \toprule
    Probe scale $\delta$
      & $0.005$ & $0.01$ & $0.02$ & $0.05$ & QP range \\
    \midrule
    Encoder $\hat L_{g_a}$ (tests $\Omega_R^{(1)}$)
      & $11.69\%$ & $12.00\%$ & $12.23\%$ & $12.24\%$ & $\phantom{0}8.5\%$--$14.8\%$ \\
    End-to-end $\hat L_f$ (tests $\Omega_D^{(1)}$)
      & $\phantom{0}8.21\%$ & $\phantom{0}8.36\%$ & $\phantom{0}8.45\%$ & $\phantom{0}8.52\%$ & $\phantom{0}7.4\%$--$10.0\%$ \\
    \bottomrule
  \end{tabular}
\end{table}

\section{Discussion and Conclusion}
\label{sec:discussion}

A single conceptual step---adding a pixel-domain Gaussian
perturbation to the encoder input on top of the standard latent
quantization proxy---yields a structured RD regularizer with a
closed-form local description: an information-weighted
encoder-Jacobian penalty (rate side) and a Frobenius contractive
penalty on the end-to-end reconstruction map (distortion side),
alongside a sign-indefinite residual--Hessian term that the rate
constraint forces to survive at finite bitrate. The two perturbations are complementary, operating at
different stages: the latent proxy provides differentiability around
quantization, while pixel-domain noise penalizes encoder sensitivity
to plausible input variation. The Jacobian analysis
(Section~\ref{sec:jacobian}) supports both predicted contractions and
preferential contraction of high-$I_\phi$ directions, while BD-Rate
gains are uniform across backbones, scales, and bitrates.

\paragraph{Connection to diffusion-style training.}
Pixel-domain noise with a clean reconstruction target gives LIC a
denoising-autoencoder interpretation: the codec still predicts the
clean image and is trained with an $x$-space loss, loosely analogous
to $x_0$-prediction in diffusion models. The optimal denoising
residual further connects this reconstruction objective to the score of
the noise-smoothed data distribution
\cite{vincent2011connection,alain2014regularized}.
This connection alone does not implement $\boldsymbol{\epsilon}$- or
$v$-prediction, which require reparameterizing the prediction target
and typically introduce noise-level-dependent loss weighting
\cite{ho2020denoising,song2021sde,karras2022edm,salimans2022progressive}.
Exploring such reparameterizations, including an $x$-prediction variant
with a $v$-space loss, is a natural direction for future work.

\paragraph{Beyond PSNR: gains on perceptual and structural metrics.}
We also fine-tune DMCI-Large with a mixed MSE,
MS-SSIM~\cite{wang2003multiscale}, LPIPS~\cite{zhang2018unreasonable},
and DISTS~\cite{ding2020image} objective. Tables~\ref{tab:perceptual_bdrate} and
\ref{tab:perceptual_vtm_bdrate} report consistent gains over both the
corresponding clean model and VTM-22.0 across all three datasets and
four metrics.

\begin{table}[t]
  \caption{Perceptual DMCI-Large: BD-Rate of noise-regularized training
           against the corresponding clean model. Negative values
           indicate bitrate savings.}
  \label{tab:perceptual_bdrate}
  \centering
  \small
  \setlength{\tabcolsep}{2.6pt}
  \begin{tabular*}{0.9\columnwidth}{@{\extracolsep{\fill}}p{0.18\columnwidth}lrrrr@{}}
    \toprule
    Dataset & PSNR & MS-SSIM & LPIPS & DISTS \\
    \midrule
    Kodak   & $-1.63\%$ & $-1.99\%$ & $-2.13\%$ & $-2.84\%$ \\
    Tecnick & $-0.63\%$ & $-0.77\%$ & $-1.12\%$ & $-1.94\%$ \\
    CLIC    & $-1.01\%$ & $-1.49\%$ & $-2.83\%$ & $-4.85\%$ \\
    \bottomrule
  \end{tabular*}
\end{table}

\begin{table}[t]
  \caption{Perceptual DMCI-Large: BD-Rate against VTM-22.0. Negative
           values indicate bitrate savings relative to VTM.}
  \label{tab:perceptual_vtm_bdrate}
  \centering
  \small
  \setlength{\tabcolsep}{2.6pt}
  \begin{tabular*}{0.9\columnwidth}{@{\extracolsep{\fill}}p{0.18\columnwidth}lrrrr@{}}
    \toprule
    Dataset & Model & PSNR & MS-SSIM & LPIPS & DISTS \\
    \midrule
    Kodak & Clean     & $-14.20\%$ & $-18.29\%$ & $-90.41\%$ & $-90.67\%$ \\
    Kodak & NoiseRegu & $\mathbf{-15.75\%}$ & $\mathbf{-20.04\%}$ & $\mathbf{-90.66\%}$ & $\mathbf{-91.01\%}$ \\
    Tecnick & Clean   & $-17.10\%$ & $-14.62\%$ & $-87.34\%$ & $-88.29\%$ \\
    Tecnick & NoiseRegu & $\mathbf{-17.65\%}$ & $\mathbf{-15.32\%}$ & $\mathbf{-87.43\%}$ & $\mathbf{-88.43\%}$ \\
    CLIC & Clean      & $-8.59\%$ & $-6.26\%$ & $-91.49\%$ & $-92.92\%$ \\
    CLIC & NoiseRegu  & $\mathbf{-9.69\%}$ & $\mathbf{-7.87\%}$ & $\mathbf{-91.65\%}$ & $\mathbf{-93.07\%}$ \\
    \bottomrule
  \end{tabular*}
\end{table}

\paragraph{Limitations.}
The theoretical analysis is a second-order expansion in $\sigma$ derived
for MSE distortion, while the perceptual results are empirical
extensions beyond this setting. We use a single exponential annealing
schedule, whose interaction with rate conditioning remains to be
studied.

\section*{Acknowledgements and Disclosures}
Funding in direct support of this work was provided by Malanshan Audio \& Video Laboratory through
institutional research and computing resources.

\bibliographystyle{unsrt}


\appendix

\section{Technical Appendices and Supplementary Material}
\label{app:technical}

This appendix collects the longer derivations that would otherwise
interrupt the main narrative of Section~\ref{sec:theory}. Its goal is to
make every algebraic step that produces the local objective
\eqref{eq:combined_regularizer} explicitly verifiable, with no missing
intermediate manipulations. The notation is kept identical to the main
text and is summarized in Table~\ref{tab:notation} for quick reference.
Throughout, all derivatives are evaluated at the clean input
$\mathbf{x}$ unless stated otherwise.

\subsection{Smoothness Assumptions}
\label{app:smoothness}
For the Gaussian-averaged Taylor expansions of
Sections~\ref{app:distortion}--\ref{app:rate} to admit an
$O(\sigma^{4})$ remainder, we assume that
$g_a$, $g_s$, and the entropy log-density $-\log p_\phi$ are
$C^{4}$ on the relevant supports of $\tilde{\mathbf{x}}$ and
$\bar{\mathbf{y}}$. The residual is then bounded by an
$O(\|\boldsymbol{\eta}\|^{4})$ Lagrange remainder, whose Gaussian
expectation is $O(\sigma^{4})$. Third-order odd moments vanish, so the
next nonzero correction after the $O(\sigma^{2})$ terms is collected in
this remainder. (A weaker $C^{3}$ assumption would yield an expected
$O(\sigma^{3})$ remainder with the same $O(\sigma^{2})$ structure.) The standard analysis and synthesis
transforms of LIC, built from convolutions with
smooth activations (e.g.\ WSiLU), satisfy this in practice.

\subsection{Clamping}
\label{app:clamping}
The expansions below use the unclamped perturbation
$\tilde{\mathbf{x}}=\mathbf{x}+\boldsymbol{\eta}$, for which the
zero-mean, isotropic Gaussian moment identities apply. Clamping can
introduce boundary corrections, but its effect diminishes as $\sigma$ is
annealed to zero. Empirically, a no-clamp DMCI-Base ablation at
$\sigma_0=0.75$ changes the average BD-Rate across Kodak, Tecnick, and
CLIC only from $-0.77\%$ with clamping to $-0.65\%$ without clamping, a
difference of $0.13$ percentage points. Thus, the noise-regularization
gain is retained without clamping.

\begin{table}[h]
\caption{Notation used throughout this appendix. Cross-references to
the main text equations are given in the right column where
applicable.}
\label{tab:notation}
\centering
\small
\renewcommand{\arraystretch}{1.15}
\begin{tabular}{@{}l p{0.62\linewidth}@{}}
\toprule
Symbol & Meaning \\
\midrule
\multicolumn{2}{@{}l}{\emph{Inputs and noise}} \\
$\mathbf{x}$ & clean input image (the distortion target) \\
$\boldsymbol{\eta}$ & encoder-side Gaussian noise, $\boldsymbol{\eta}\!\sim\!\mathcal{N}(0,\sigma^2 I)$ \\
$\sigma$ & noise scale (the small parameter of the local expansion) \\
$\tilde{\mathbf{x}}=\mathbf{x}+\boldsymbol{\eta}$ & noisy encoder input \\
$\mathbf{u}$ & uniform proxy for latent quantization, $\mathbf{u}\!\sim\!\mathcal{U}(-\tfrac{1}{2},\tfrac{1}{2})$ \\
\midrule
\multicolumn{2}{@{}l}{\emph{Codec, Jacobians, and Hessians}} \\
$g_a,\,g_s$ & analysis (encoder) and synthesis (decoder) transforms \\
$f(\mathbf{x})=g_s\bigl(g_a(\mathbf{x})+\mathbf{u}\bigr)$ & training-time end-to-end reconstruction map (fixed $\mathbf{u}$) \\
$J_{g_a}(\mathbf{x}),\,J_{f}(\mathbf{x})$ & Jacobian of $g_a$ and $f$ at $\mathbf{x}$ \\
$\nabla^2 f_k(\mathbf{x})$ & Hessian of the scalar coordinate $f_k:\mathbb{R}^d\!\to\!\mathbb{R}$ at $\mathbf{x}$ \\
$H_{g_a}(\mathbf{x})$ & component-wise Hessian tensor of $g_a$, see Eq.~\eqref{eq:dy_expansion} \\
$\Delta f(\mathbf{x}),\,\Delta g_a(\mathbf{x})$ & component-wise Laplacian of $f$ and $g_a$ at $\mathbf{x}$ \\
\midrule
\multicolumn{2}{@{}l}{\emph{Latent and entropy model}} \\
$\bar{\mathbf{y}}_0=g_a(\mathbf{x})+\mathbf{u}$ & training-time latent at the clean input \\
$\bar{\mathbf{y}}=g_a(\tilde{\mathbf{x}})+\mathbf{u}$ & training-time latent at the noisy input \\
$\delta\mathbf{y}=\bar{\mathbf{y}}-\bar{\mathbf{y}}_0$ & latent perturbation induced by $\boldsymbol{\eta}$ \\
$p_\phi$ & entropy model on the latent $\mathbf{y}$ \\
$s_\phi(\mathbf{y})=\nabla\log p_\phi(\mathbf{y})$ & score of the entropy model \\
$I_\phi(\mathbf{y})=-\nabla^2\log p_\phi(\mathbf{y})$ & observed information matrix of $p_\phi$ \\
\midrule
\multicolumn{2}{@{}l}{\emph{Loss decomposition}} \\
$D(\tilde{\mathbf{x}},\mathbf{x})=\|f(\tilde{\mathbf{x}})-\mathbf{x}\|_2^2$ & per-sample distortion (clean target) \\
$R(\tilde{\mathbf{x}})=-\log p_\phi(\bar{\mathbf{y}})$ & per-sample rate \\
$\mathcal{L}_0,\,\mathcal{L}_{\mathrm{noise}}$ & clean-input and noisy-input RD losses, Eq.~\eqref{eq:noise_loss} \\
$\Omega_R^{(1)}$ & information-weighted Jacobian penalty $\mathrm{tr}(J_{g_a}^{\!\top} I_\phi J_{g_a})$ \\
$\Omega_R^{(2)}$ & rate-side score--Laplacian term $-s_\phi^{\!\top}\Delta g_a$ \\
$\Omega_D^{(1)}$ & end-to-end Frobenius penalty $\|J_f\|_F^{2}$ \\
$\Omega_D^{(2)}$ & residual--Laplacian term $\langle f-\mathbf{x},\,\Delta f\rangle$ \\
$\lambda$ & RD trade-off coefficient \\
\bottomrule
\end{tabular}
\end{table}

\subsection{Distortion-Side Expansion}
\label{app:distortion}

This subsection derives the $O(\sigma^2)$ correction to the distortion
branch announced in Eq.~\eqref{eq:distortion_expansion_main} of the
main text. We work in five elementary steps.

\paragraph{Setup.}
For a fixed quantization proxy $\mathbf{u}$, define the training-time
end-to-end reconstruction map and the per-sample distortion as
\begin{equation}
  f(\mathbf{x}) \;:=\; g_s\bigl(g_a(\mathbf{x})+\mathbf{u}\bigr),
  \qquad
  D(\tilde{\mathbf{x}},\mathbf{x}) \;:=\; \|f(\tilde{\mathbf{x}})-\mathbf{x}\|_2^2.
  \label{eq:f_def}
\end{equation}
Because the distortion target is the clean image $\mathbf{x}$, the
noise $\boldsymbol{\eta}$ enters $D$ \emph{only} through the argument
of $f$; this is the structural reason why the distortion branch
produces both a Jacobian penalty and a residual--Hessian term.

\paragraph{Step 1: Taylor expansion of each output coordinate.}
For any output coordinate $f_k$ of the map $f$, a second-order Taylor
expansion of $f_k$ around the clean input $\mathbf{x}$ gives
\begin{equation}
  f_k(\tilde{\mathbf{x}})
  \;=\; f_k(\mathbf{x})
  \;+\; \nabla f_k(\mathbf{x})^{\!\top}\boldsymbol{\eta}
  \;+\; \tfrac{1}{2}\,\boldsymbol{\eta}^{\!\top}\nabla^2 f_k(\mathbf{x})\,\boldsymbol{\eta}
  \;+\; O(\|\boldsymbol{\eta}\|^3),
  \label{eq:taylor_f}
\end{equation}
where $\nabla f_k$ and $\nabla^2 f_k$ denote the gradient and Hessian
of the scalar function $f_k:\mathbb{R}^d\!\to\!\mathbb{R}$. From now
on, derivatives are evaluated at $\mathbf{x}$ unless stated otherwise.

\paragraph{Step 2: Squaring and grouping by order.}
Substituting Eq.~\eqref{eq:taylor_f} into the squared error,
expanding the bracket, and collecting terms by their order in
$\boldsymbol{\eta}$, we obtain
\begin{equation}
\begin{aligned}
  \bigl\|f(\tilde{\mathbf{x}})-\mathbf{x}\bigr\|_2^{2}
  \;=\;& \sum_{k}\bigl(f_k(\mathbf{x})-x_k\bigr)^{2}
  \\
  &+\; 2\sum_{k}\bigl(f_k(\mathbf{x})-x_k\bigr)\,\nabla f_k^{\!\top}\boldsymbol{\eta}
  \\
  &+\; \sum_{k}\bigl(f_k(\mathbf{x})-x_k\bigr)\,\boldsymbol{\eta}^{\!\top}\nabla^2 f_k\,\boldsymbol{\eta}
  \\
  &+\; \sum_{k}\bigl(\nabla f_k^{\!\top}\boldsymbol{\eta}\bigr)^{2}
  \;+\; O\bigl(\|\boldsymbol{\eta}\|^{3}\bigr).
\end{aligned}
  \label{eq:dist_expanded}
\end{equation}
The four lines on the right-hand side are, in order, the clean-input
distortion $D_0(\mathbf{x})=\|f(\mathbf{x})-\mathbf{x}\|_2^{2}$, a
linear-in-$\boldsymbol{\eta}$ term, a residual$\times$Hessian
quadratic, and a squared-Jacobian quadratic.

\paragraph{Step 3: Gaussian moment identities.}
For $\boldsymbol{\eta}\!\sim\!\mathcal{N}(0,\sigma^2 I)$, the first
two moments are $\mathbb{E}[\boldsymbol{\eta}]=\mathbf{0}$ and
$\mathbb{E}[\boldsymbol{\eta}\boldsymbol{\eta}^{\!\top}]=\sigma^2 I$.
These imply, for any deterministic vector
$\mathbf{a}\in\mathbb{R}^{d}$ and any deterministic symmetric matrix
$A\in\mathbb{R}^{d\times d}$, the elementary contractions
\begin{equation}
  \mathbb{E}\bigl[\mathbf{a}^{\!\top}\boldsymbol{\eta}\bigr] \;=\; 0,
  \qquad
  \mathbb{E}\bigl[(\mathbf{a}^{\!\top}\boldsymbol{\eta})^2\bigr] \;=\; \sigma^2\,\|\mathbf{a}\|_2^2,
  \qquad
  \mathbb{E}\bigl[\boldsymbol{\eta}^{\!\top}\!A\,\boldsymbol{\eta}\bigr] \;=\; \sigma^2\,\mathrm{tr}(A).
  \label{eq:gauss_quadratic}
\end{equation}
These three identities are the workhorses of every step that follows
in both Section~\ref{app:distortion} and Section~\ref{app:rate}.

\paragraph{Step 4: Term-by-term expectation.}
Apply Eq.~\eqref{eq:gauss_quadratic} to each line of
Eq.~\eqref{eq:dist_expanded}:
\begin{itemize}
  \item the linear-in-$\boldsymbol{\eta}$ term has zero expectation;
  \item the residual$\times$Hessian term contracts via
        $\mathbb{E}\bigl[\boldsymbol{\eta}^{\!\top}\nabla^2 f_k\,\boldsymbol{\eta}\bigr]
        =\sigma^2\,\mathrm{tr}\bigl(\nabla^2 f_k\bigr)
        =\sigma^2\,\Delta f_k(\mathbf{x})$;
  \item the squared-Jacobian term contracts via
        $\mathbb{E}\bigl[(\nabla f_k^{\!\top}\boldsymbol{\eta})^2\bigr]
        =\sigma^2\,\|\nabla f_k\|_2^{2}$.
\end{itemize}
Summing over the output index $k$, the total $O(\sigma^2)$
contribution to the expected distortion becomes
\begin{equation}
  \sigma^2\sum_{k}\bigl(f_k(\mathbf{x})-x_k\bigr)\,\Delta f_k(\mathbf{x})
  \;+\;
  \sigma^2\sum_{k}\bigl\|\nabla f_k(\mathbf{x})\bigr\|_2^{2}.
\end{equation}

\paragraph{Step 5: Compact form.}
The two sums above are, respectively, the residual--Laplacian inner
product and the squared Frobenius norm of the Jacobian:
\begin{equation}
  \sum_{k}\bigl(f_k(\mathbf{x})-x_k\bigr)\,\Delta f_k(\mathbf{x})
  \;=\; \bigl\langle f(\mathbf{x})-\mathbf{x},\,\Delta f(\mathbf{x})\bigr\rangle,
  \qquad
  \sum_{k}\bigl\|\nabla f_k(\mathbf{x})\bigr\|_2^{2}
  \;=\; \|J_f(\mathbf{x})\|_F^{2}.
  \label{eq:dist_compact_id}
\end{equation}
Combining the clean-input contribution with the two $O(\sigma^2)$
corrections yields the distortion-side expansion in the form quoted
as Eq.~\eqref{eq:distortion_expansion_main} of the main text:
\begin{equation}
\begin{aligned}
  \mathbb{E}_{\boldsymbol{\eta}}\bigl[D(\tilde{\mathbf{x}},\mathbf{x})\bigr]
  \;=\;& \underbrace{\|f(\mathbf{x})-\mathbf{x}\|_2^{2}}_{D_0(\mathbf{x})} \\
  +\;& \sigma^2\,\underbrace{\|J_f(\mathbf{x})\|_F^{2}}_{\Omega_D^{(1)}} \\
  +\;& \sigma^2\,\underbrace{\bigl\langle f(\mathbf{x})-\mathbf{x},\,\Delta f(\mathbf{x})\bigr\rangle}_{\Omega_D^{(2)}}
  \;+\; O(\sigma^4).
\end{aligned}
  \label{eq:distortion_app}
\end{equation}
The first $O(\sigma^2)$ correction $\sigma^2\,\Omega_D^{(1)}$ is the
end-to-end contractive Jacobian penalty discussed in
Section~\ref{sec:dist_branch}. The second, $\sigma^2\,\Omega_D^{(2)}$, is
the LIC-specific residual--Hessian term: it survives at $O(\sigma^2)$
precisely because the rate constraint keeps
$f(\mathbf{x})\!\neq\!\mathbf{x}$ at finite bitrate, whereas a
classical denoising autoencoder satisfies
$f^\star(\mathbf{x})\!\approx\!\mathbf{x}$ at the optimum and the same
term is suppressed to $O(\sigma^4)$.

\paragraph{Equivalent Laplacian form.}
Differentiating the squared error $\sum_{k}(f_k(\mathbf{x})-x_k)^{2}$
twice with respect to $\mathbf{x}$ and contracting indices yields the
pointwise identity
\begin{equation}
  \tfrac{1}{2}\,\Delta\bigl\|f(\mathbf{x})-\mathbf{x}\bigr\|_2^{2}
  \;=\; \bigl\|J_f(\mathbf{x})-I\bigr\|_F^{2}
  \;+\; \bigl\langle f(\mathbf{x})-\mathbf{x},\,\Delta f(\mathbf{x})\bigr\rangle,
  \label{eq:lap_id}
\end{equation}
which can be substituted into Eq.~\eqref{eq:distortion_app} to obtain
an equivalent decomposition of the $O(\sigma^{2})$ distortion correction
in terms of $\|J_f\|_F^{2}$, $\|J_f-I\|_F^{2}$, and
$\Delta\|f-\mathbf{x}\|_2^{2}$. The two formulations agree to leading
order in $\sigma$.

\subsection{Rate-Side Expansion}
\label{app:rate}

This subsection derives the $O(\sigma^{2})$ correction to the rate
branch announced as Eq.~\eqref{eq:rate_expansion_main}, in three
logical movements: expand $g_a$ in $\boldsymbol{\eta}$ to obtain the
latent perturbation $\delta\mathbf{y}$; read off its first two
moments; and propagate them through a second-order expansion of
$-\log p_\phi$. All derivatives are evaluated at the clean point
($\mathbf{x}$ for $g_a$ and its derivatives, $\bar{\mathbf{y}}_0$ for
$s_\phi$ and $I_\phi$) unless stated otherwise.

\paragraph{Step 1: Latent perturbation from a noisy input.}
Taylor-expanding $g_a$ around $\mathbf{x}$ to second order in
$\boldsymbol{\eta}$ and subtracting
$\bar{\mathbf{y}}_0 = g_a(\mathbf{x})+\mathbf{u}$ from both sides of
$\bar{\mathbf{y}} = g_a(\tilde{\mathbf{x}})+\mathbf{u}$, the proxy
$\mathbf{u}$ cancels and we obtain
\begin{equation}
  \delta\mathbf{y}
  \;=\; J_{g_a}\,\boldsymbol{\eta}
  \;+\; \tfrac{1}{2}\,H_{g_a}[\boldsymbol{\eta},\boldsymbol{\eta}]
  \;+\; O\bigl(\|\boldsymbol{\eta}\|^{3}\bigr),
  \qquad
  \bigl(H_{g_a}[\boldsymbol{\eta},\boldsymbol{\eta}]\bigr)_{k}
  := \boldsymbol{\eta}^{\!\top}\nabla^{2}(g_a)_{k}\,\boldsymbol{\eta},
  \label{eq:dy_expansion}
\end{equation}
where the component-wise contraction is indexed by the output
coordinate~$k$.

\paragraph{Step 2: First two moments of $\delta\mathbf{y}$.}
Applying the Gaussian identities~\eqref{eq:gauss_quadratic}
component-wise to Eq.~\eqref{eq:dy_expansion}: the linear term has
zero mean; the Hessian term contributes to the mean at $O(\sigma^{2})$
and to the covariance only at $O(\sigma^{4})$. We read off
\begin{equation}
  \mathbb{E}[\delta\mathbf{y}]
  \;=\; \tfrac{\sigma^{2}}{2}\,\Delta g_a,
  \qquad
  \mathbb{E}[\delta\mathbf{y}\,\delta\mathbf{y}^{\!\top}]
  \;=\; \sigma^{2}\,J_{g_a}\,J_{g_a}^{\!\top} + O(\sigma^{4}).
  \label{eq:dy_moments}
\end{equation}

\paragraph{Step 3: Local expansion of the rate and its expectation.}
A second-order expansion of $R(\tilde{\mathbf{x}}) = -\log p_\phi(\bar{\mathbf{y}})$
around $\bar{\mathbf{y}}_0$ in $\delta\mathbf{y}$ gives, with
$\nabla(-\log p_\phi)=-s_\phi$ and $\nabla^{2}(-\log p_\phi)=I_\phi$,
\begin{equation}
  R(\tilde{\mathbf{x}})
  \;=\; R_0(\mathbf{x})
  \;-\; s_\phi^{\!\top}\delta\mathbf{y}
  \;+\; \tfrac{1}{2}\,\delta\mathbf{y}^{\!\top} I_\phi\,\delta\mathbf{y}
  \;+\; O\bigl(\|\delta\mathbf{y}\|^{3}\bigr),
  \label{eq:rate_local_expansion}
\end{equation}
where $R_0(\mathbf{x}) := -\log p_\phi(\bar{\mathbf{y}}_0)$ is the
clean-input rate. Taking the expectation over $\boldsymbol{\eta}$ and
substituting Eq.~\eqref{eq:dy_moments} (using
$\mathbb{E}[\delta\mathbf{y}^{\!\top}I_\phi\,\delta\mathbf{y}]
= \mathrm{tr}(I_\phi\,\mathbb{E}[\delta\mathbf{y}\delta\mathbf{y}^{\!\top}])$
and the trace's cyclicity) yields the rate-side expansion in the form
quoted as Eq.~\eqref{eq:rate_expansion_main} of the main text,
\begin{equation}
  \mathbb{E}_{\boldsymbol{\eta}}\bigl[R(\tilde{\mathbf{x}})\bigr]
  \;=\; R_0(\mathbf{x})
  \;+\; \tfrac{\sigma^{2}}{2}\,\Omega_R^{(1)}(\mathbf{x})
  \;+\; \tfrac{\sigma^{2}}{2}\,\Omega_R^{(2)}(\mathbf{x})
  \;+\; O(\sigma^{4}),
  \label{eq:rate_app}
\end{equation}
with the two rate-side $O(\sigma^{2})$ corrections
\begin{equation}
  \Omega_R^{(1)}(\mathbf{x})
  := \mathrm{tr}\!\bigl(J_{g_a}^{\!\top}\,I_\phi\,J_{g_a}\bigr),
  \qquad
  \Omega_R^{(2)}(\mathbf{x})
  := -\,s_\phi^{\!\top}\,\Delta g_a.
  \label{eq:OmegaR_app}
\end{equation}
Cross terms with the latent proxy $\mathbf{u}$ vanish at this order:
$\mathbf{u}$ is independent of $\boldsymbol{\eta}$ and enters only
through $\bar{\mathbf{y}}_0$, contributing no
$\boldsymbol{\eta}$-dependent term up to $O(\sigma^{4})$. The
interpretation of the two corrections is taken up jointly with the
distortion-side terms in Appendix~\ref{app:summary}.

\paragraph{Hyperprior and context models.}
For a hyperprior, the perturbation also propagates through
$\bar{\mathbf{z}}=h_a(\bar{\mathbf{y}})$, adding a curvature term
\[
  \Omega_z=\mathrm{tr}\!\left(J_{h_a\circ g_a}^{\top} I_z
  J_{h_a\circ g_a}\right),
\]
together with cross-Hessian and drift terms. The curvature term is
positive semidefinite for Gaussian-type hyperpriors; the remaining terms
need not have a definite sign. Context models further introduce
off-diagonal information entries, so the diagonal inverse-variance
reading of $I_\phi$ applies only to elementwise mean-scale Gaussian
conditionals.

\subsection{Combined Local Objective}
\label{app:summary}

We now assemble the distortion-side expansion \eqref{eq:distortion_app}
and the rate-side expansion \eqref{eq:rate_app} into the local form
of the noisy RD loss quoted as Eq.~\eqref{eq:combined_regularizer}.
Multiplying Eq.~\eqref{eq:distortion_app} by $\lambda$, adding it to
Eq.~\eqref{eq:rate_app}, and grouping each $O(\sigma^{2})$ correction
with its branch yields
\begin{equation}
\begin{aligned}
  \mathcal{L}_{\mathrm{noise}}(\mathbf{x})
  \;=\;& \mathcal{L}_0(\mathbf{x})
  \;+\; \tfrac{\sigma^{2}}{2}\,\bigl(\Omega_R^{(1)}(\mathbf{x}) + \Omega_R^{(2)}(\mathbf{x})\bigr)
  \\
  \;+\;& \lambda\,\sigma^{2}\,\bigl(\Omega_D^{(1)}(\mathbf{x}) + \Omega_D^{(2)}(\mathbf{x})\bigr)
  \;+\; O(\sigma^{4}),
\end{aligned}
  \label{eq:combined_app}
\end{equation}
with $\mathcal{L}_0(\mathbf{x}) := R_0(\mathbf{x}) + \lambda\,D_0(\mathbf{x})$
the clean-input RD loss, the rate-side corrections
$\Omega_R^{(1)},\Omega_R^{(2)}$ given by
Eq.~\eqref{eq:OmegaR_app}, and the distortion-side corrections
$\Omega_D^{(1)},\Omega_D^{(2)}$ given by
Eq.~\eqref{eq:distortion_app}. Eq.~\eqref{eq:combined_app} is the
local objective of the main text
(Eq.~\eqref{eq:combined_regularizer}) in expanded form. The four
corrections have distinct status:
$\Omega_R^{(1)}$ is the sign-definite, information-aware encoder
contraction;
$\Omega_R^{(2)}$ is the rate-side score--Laplacian term, locally
nonzero and sign-indefinite per sample, with a marginal expectation
that is not strictly zero in general but exhibits zero-mean
oscillatory behavior at a well-calibrated entropy model;
$\Omega_D^{(1)}=\|J_f\|_F^{2}$ is the sign-definite, end-to-end codec
contraction, measured directly in Section~\ref{sec:jacobian};
$\Omega_D^{(2)}$ is the LIC-specific, sign-indefinite
residual--Laplacian burden. The dynamic role of $\sigma$ in trading
off the useful pair $\{\Omega_R^{(1)},\,\Omega_D^{(1)}\}$ against the
LIC-specific cost $\Omega_D^{(2)}$ is taken up in
Appendix~\ref{app:annealing}.

\subsection{Connections to Information Bottleneck and Score Matching}
\label{app:connections}

This subsection expands the two perspectives sketched at the end of
Section~\ref{sec:combined}. The full derivations are not strictly
needed to use the local objective \eqref{eq:combined_regularizer},
but they make explicit the connections to two well-known principles.

\paragraph{Information-bottleneck connection.}
With $\tilde{\mathbf{x}}=\mathbf{x}+\boldsymbol{\eta}$ and a clean
distortion target $\mathbf{x}$, the rate term upper-bounds
$I(\mathbf{y};\tilde{\mathbf{x}})$, and the squared-error distortion
is, up to a data entropy that does not depend on the encoder, an
upper bound on $H(\mathbf{x}\!\mid\!\mathbf{y})$, so its negation
\emph{lower}-bounds $I(\mathbf{y};\mathbf{x})$. The data-processing
inequality gives
$I(\mathbf{y};\mathbf{x})\le I(\mathbf{y};\tilde{\mathbf{x}})$, with
equality iff $\mathbf{y}$ is conditionally independent of
$\boldsymbol{\eta}$ given $\mathbf{x}$. Minimizing
$\mathcal{L}_{\mathrm{noise}}$ therefore pushes the encoder toward
the conditional-independence optimum: the latent retains image
content and discards $\boldsymbol{\eta}$. This is mechanistically
consistent with $\Omega_R^{(1)}$, which shrinks the encoder Jacobian
precisely along the expensive directions through which
$\boldsymbol{\eta}$ would otherwise leak into the latent.

\paragraph{Score-matching connection.}
If we drop the rate term and focus only on the noisy-input,
clean-target distortion, classical DAE
theory~\cite{vincent2011connection,alain2014regularized} gives the
optimum
\begin{equation}
  f^{\star}(\tilde{\mathbf{x}}) - \tilde{\mathbf{x}}
  \;=\; \sigma^{2}\,\nabla_{\tilde{\mathbf{x}}}\log q_\sigma(\tilde{\mathbf{x}}),
  \qquad
  q_\sigma = p_{\mathrm{data}}\!\ast\!\mathcal{N}(0,\sigma^{2}I),
  \label{eq:dae_optimum}
\end{equation}
i.e.\ the optimal residual aligns at $O(\sigma^{2})$ with the Stein
score of the noise-smoothed data distribution, biasing the codec
toward high-density image regions. In LIC the rate constraint
perturbs this optimal vector field at $O(\sigma^{2})$, and the
resulting shift in the stationary point of the loss is governed by
the inclusion of the LIC-specific scalar penalty $\Omega_D^{(2)}$
in Eq.~\eqref{eq:combined_regularizer}.

\subsection{Why Annealing Matters}
\label{app:annealing}

The local objective \eqref{eq:combined_regularizer} shows that every
$O(\sigma^{2})$ correction carries an explicit factor of $\sigma^{2}$;
so does the gradient field,
\begin{equation}
  \nabla_\theta \mathcal{L}(\theta;\sigma(t))
  \;\approx\;
  \nabla_\theta \mathcal{L}_0(\theta)
  \;+\;
  \sigma(t)^{2}\,\nabla_\theta \Omega(\theta),
  \label{eq:annealed_grad}
\end{equation}
with $\Omega$ the combined regularizer of
Eq.~\eqref{eq:combined_regularizer}. Because $\sigma$ multiplies all
four corrections jointly, scheduling it modulates the regularization
\emph{as a whole}; what changes between training phases is therefore
not the relative weighting between $\Omega_R^{(1)},\Omega_D^{(1)}$
and $\Omega_D^{(2)}$, but how much the entire $\Omega$ is allowed to
deflect the gradient field away from the clean RD loss
$\nabla_\theta\mathcal{L}_0$.

\paragraph{Why anneal at all (and not just train at fixed $\sigma$).}
A fixed-$\sigma$ run pays a permanent $O(\sigma^{2})$ correction from
all four terms at convergence; among them, the sign-indefinite
residual--Hessian cost $\Omega_D^{(2)}$ is the dominant remaining
penalty at low rate, since the rate constraint forces
$f-\mathbf{x}\!\neq\!0$ regardless of how well-trained the network is.
Annealing instead separates the regularizer's two jobs in time:
(i) \emph{geometric shaping}, front-loaded.\ Early in training the
network is still searching for a manifold parameterization, and a
large $\sigma$ activates the helpful pair
$(\Omega_R^{(1)},\Omega_D^{(1)})$ as a strong prior that contracts
both encoder and end-to-end Jacobians along expensive directions.
(ii) \emph{Rate--distortion fine-tuning}, back-loaded.\ Late in
training the manifold structure has been mostly captured, and any
residual $O(\sigma^{2})$ correction acts as a constraint rather
than a guide; sending $\sigma(t)\!\to\!0$ removes \emph{all four}
corrections together, so the helpful pair is shed alongside
$\Omega_D^{(2)}$, but by this point the helpful pair has already
done its shaping work, while the harmful term would otherwise
persist throughout training. The optimizer then completes
convergence on the clean RD loss, where it can fine-tune the
rate--distortion trade-off precisely.

\paragraph{Coupling with entropy-model maturation.}
The information-weighted Jacobian penalty $\Omega_R^{(1)}$ is only as
informative as the entropy model that supplies $I_\phi$. Early in
training $I_\phi$ tracks an immature density estimator and the penalty
is correspondingly noisy; late in training $I_\phi$ is accurate but
$\sigma^{2}(t)$ is small. The two effects pull in opposite temporal
directions, so a sensible schedule keeps $\sigma$ moderate during the
window in which the entropy model matures, rather than front-loading
all of the noise. In practice $\sigma_0\!\in\![0.1,0.5]$ on the
$[0,1]$ pixel scale, an annealing horizon overlapping the entropy
model's calibration period, and a terminal $\sigma\!\approx\!0$
jointly satisfy these requirements; this is the design behind
Eq.~\eqref{eq:schedule}, and it is empirically corroborated by the
$\sigma_0$ sweep in Table~\ref{tab:sigma_ablation}, where any
sufficiently large initial scale ($\sigma_0\!\gtrsim\!0.5$) saturates
the BD-Rate gain on DMCI-Base.

\subsection{Drop-in Training Interface}
\label{app:lic_noise}

\lstdefinestyle{pythoncompact}{%
  language=Python,
  basicstyle=\ttfamily\footnotesize,
  keywordstyle=\bfseries\color{blue!60!black},
  commentstyle=\itshape\color{gray!70!black},
  stringstyle=\color{purple!60!black},
  showstringspaces=false,
  breaklines=true,
  columns=fullflexible,
  numbers=none,
  xleftmargin=0pt,
  xrightmargin=0pt,
  aboveskip=2pt,
  belowskip=2pt,
}
Listing~\ref{lst:lic_noise} gives a minimal PyTorch sketch of the
training loop in Algorithm~\ref{alg:nat}. The variables \texttt{x},
\texttt{x\_tilde}, \texttt{x\_hat}, and \texttt{sigma\_batch}
correspond to $\mathbf{x}$, $\tilde{\mathbf{x}}$, $\hat{\mathbf{x}}$,
and $\sigma_{\mathrm{batch}}$, respectively; the schedule in
Eq.~\eqref{eq:schedule} is wrapped by \texttt{lic\_noise}.

\begin{lstlisting}[style=pythoncompact,float=t,label={lst:lic_noise},
                   caption={Minimal PyTorch sketch of the training loop.
                     The callable \texttt{lic\_noise} applies the
                     schedule in Eq.~\ref{eq:schedule} and injects the
                     Gaussian perturbation of Algorithm~\ref{alg:nat}.}]
from lic_noise import lic_noise

noise = lic_noise(noise_max=sigma_0, noise_epoch=T, total_epoch=T_total)

for epoch in range(T_total):
    for x in dataloader:
        x_tilde, sigma_batch = noise(x, epoch)
        x_hat, rate = codec(x_tilde)
        loss = rate + lam * mse(x_hat, x)
        optimizer.zero_grad()
        loss.backward()
        optimizer.step()
\end{lstlisting}

\subsection{Full Lipschitz Sweep Tables}
\label{app:lipschitz_full}

For completeness we report the full per-QP, per-$\delta$ relative
reductions underlying the summary in
Table~\ref{tab:lipschitz_summary} of the main text.
Table~\ref{tab:lipschitz_sweep_full_enc} covers the encoder local
Lipschitz constant $\hat L_{g_a}$ ($\Phi=g_a$ in
Eq.~\eqref{eq:lip_estimate}), the empirical signature of the rate-side
penalty $\Omega_R^{(1)}$.
Table~\ref{tab:lipschitz_sweep_full_codec} covers the end-to-end codec
Lipschitz constant $\hat L_f$ ($\Phi=f$), the empirical signature of
the distortion-side penalty $\Omega_D^{(1)}=\|J_f\|_F^{2}$.

\begin{table}[h]
  \centering
  \small
  \setlength{\tabcolsep}{5pt}
  \caption{%
    \textbf{Full sweep: relative reduction (\%) in encoder local
    Lipschitz $\hat{L}_{g_a}$ on Kodak (DMCI-Base, baseline $\to$
    noise-regularized).} Each cell averages
    Eq.~\eqref{eq:lip_estimate} ($\Phi=g_a$) over $24$ Kodak images
    and $n\!=\!64$ random directions. Row averages stay within a
    $0.6$-pp band as $\delta$ varies over an order of magnitude.}
  \label{tab:lipschitz_sweep_full_enc}
  \begin{tabular}{c|cccccccc|c}
    \toprule
    $\delta$ & QP$\,=\,$0 & 9 & 18 & 27 & 36 & 45 & 54 & 63 & avg.\ \\
    \midrule
    $0.005$ & 10.28 & \phantom{0}9.33 & 8.54 & 10.17 & 11.99 & 13.71 & 14.82 & 14.70 & 11.69 \\
    $0.01$  & 11.14 & 10.06           & 9.22 & 10.42 & 12.25 & 13.60 & 14.68 & 14.61 & 12.00 \\
    $0.02$  & 11.91 & 10.76           & 9.71 & 10.59 & 12.73 & 13.32 & 14.42 & 14.39 & 12.23 \\
    $0.05$  & 12.90 & 11.49           & 9.85 & 10.27 & 12.42 & 12.75 & 14.03 & 14.22 & 12.24 \\
    \bottomrule
  \end{tabular}
\end{table}

\begin{table}[h]
  \centering
  \small
  \setlength{\tabcolsep}{5pt}
  \caption{%
    \textbf{Full sweep: relative reduction (\%) in end-to-end codec
    local Lipschitz $\hat{L}_f$ on Kodak (DMCI-Base, baseline $\to$
    noise-regularized).} Each cell averages
    Eq.~\eqref{eq:lip_estimate} applied to $\Phi=f$ over $24$ Kodak
    images and $n\!=\!64$ random directions. Row averages stay within
    a $0.31$-pp band; the per-QP trend is opposite to that of the
    encoder, with the strongest reduction at the lowest rate.}
  \label{tab:lipschitz_sweep_full_codec}
  \begin{tabular}{c|cccccccc|c}
    \toprule
    $\delta$ & QP$\,=\,$0 & 9 & 18 & 27 & 36 & 45 & 54 & 63 & avg.\ \\
    \midrule
    $0.005$ & 9.33 & 8.84 & 8.36 & 8.75 & 7.51 & 7.82 & 7.44 & 7.62 & 8.21 \\
    $0.01$  & 9.56 & 9.10 & 8.46 & 8.72 & 7.64 & 7.97 & 7.56 & 7.85 & 8.36 \\
    $0.02$  & 9.81 & 9.30 & 8.47 & 8.79 & 7.83 & 8.02 & 7.47 & 7.86 & 8.45 \\
    $0.05$  & 9.99 & 9.57 & 8.45 & 8.56 & 7.98 & 8.20 & 7.38 & 8.05 & 8.52 \\
    \bottomrule
  \end{tabular}
\end{table}

\subsection{Additional Backbone: HPCM-Large}
\label{app:hpcm}

For completeness, we also apply the noise-regularized training to
HPCM-Large~\cite{li2025hpcm} as a preliminary out-of-sample check
beyond the headline ELIC/DMCI evaluations. On this auxiliary backbone
we evaluate at a single calibrated operating point per dataset; a
multi-rate sweep that would yield a proper BD-Rate is left to
follow-up work. Table~\ref{tab:hpcm_large_app} reports paired
bpp/PSNR numbers for the reproduced clean-trained baseline and the
noise-regularized variant on Kodak, CLIC, and Tecnick. At each anchor,
the noise-regularized model lowers bitrate by $3.7$--$4.6\%$ with only
a small PSNR drop ($-0.07$ to $-0.12$\,dB), a trend in the same
direction as the rate--distortion improvements observed on ELIC and
DMCI in Table~\ref{tab:bdrate_results}. While a single operating
point per dataset does not allow us to compute a BD-Rate, these
measurements are consistent with a favorable transfer of the recipe
to this third architecture; the full multi-rate verification is
deferred to follow-up work.

\begin{table}[h]
\caption{HPCM-Large at a single calibrated operating point on Kodak,
CLIC, and Tecnick. We list bpp and PSNR for the reproduced
clean-trained baseline and the noise-regularized variant; $\Delta$bpp
is the relative bitrate change and $\Delta$PSNR the absolute quality
change of the noise-regularized variant relative to the corresponding
baseline. We do not aggregate these single-point pairs into a BD-Rate.}
\label{tab:hpcm_large_app}
\centering
\small
\setlength{\tabcolsep}{6pt}
\begin{tabular}{@{}llccrr@{}}
\toprule
Dataset & Variant & bpp & PSNR (dB) & $\Delta$bpp & $\Delta$PSNR (dB) \\
\midrule
\multirow{2}{*}{Kodak}
  & HPCM-Large (baseline)         & 0.8313 & 38.022 & ---       & ---     \\
  & HPCM-Large w/ Noise Reg.\     & 0.7965 & 37.906 & $-4.19\%$ & $-0.116$ \\
\midrule
\multirow{2}{*}{CLIC}
  & HPCM-Large (baseline)         & 0.6151 & 38.691 & ---       & ---     \\
  & HPCM-Large w/ Noise Reg.\     & 0.5867 & 38.598 & $-4.61\%$ & $-0.093$ \\
\midrule
\multirow{2}{*}{Tecnick}
  & HPCM-Large (baseline)         & 0.5613 & 38.667 & ---       & ---     \\
  & HPCM-Large w/ Noise Reg.\     & 0.5403 & 38.598 & $-3.74\%$ & $-0.070$ \\
\bottomrule
\end{tabular}
\end{table}

\subsection{Per-Dataset RD Curves}
\label{app:rd_curves}

\begin{figure}[p]
  \centering
  \includegraphics[width=\linewidth]{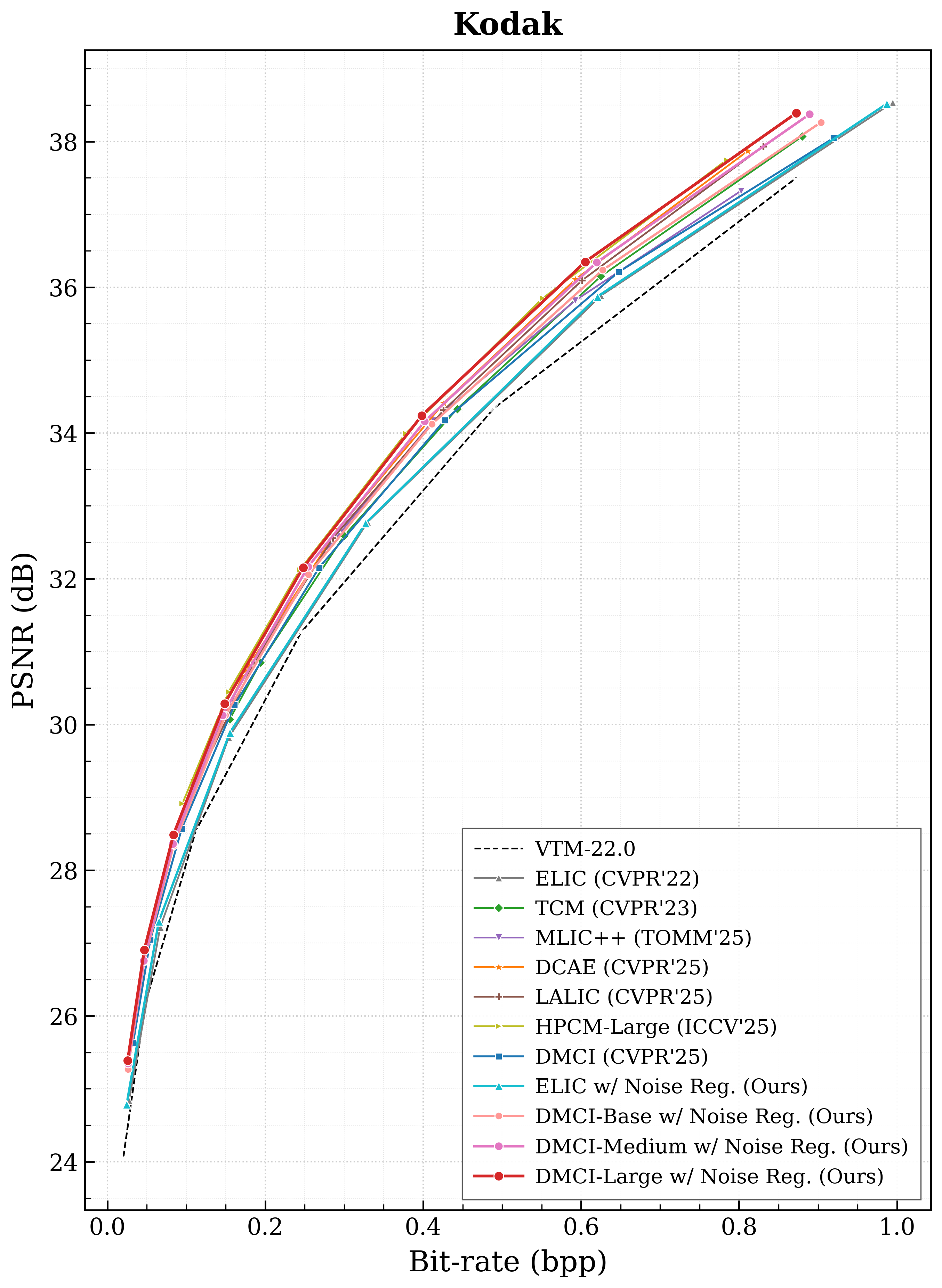}
  \caption{RD curves on \textbf{Kodak}. PSNR vs.\ bitrate (bpp).
           Curves closer to the
           upper-left corner are better.}
  \label{fig:rd_kodak}
\end{figure}

\begin{figure}[p]
  \centering
  \includegraphics[width=\linewidth]{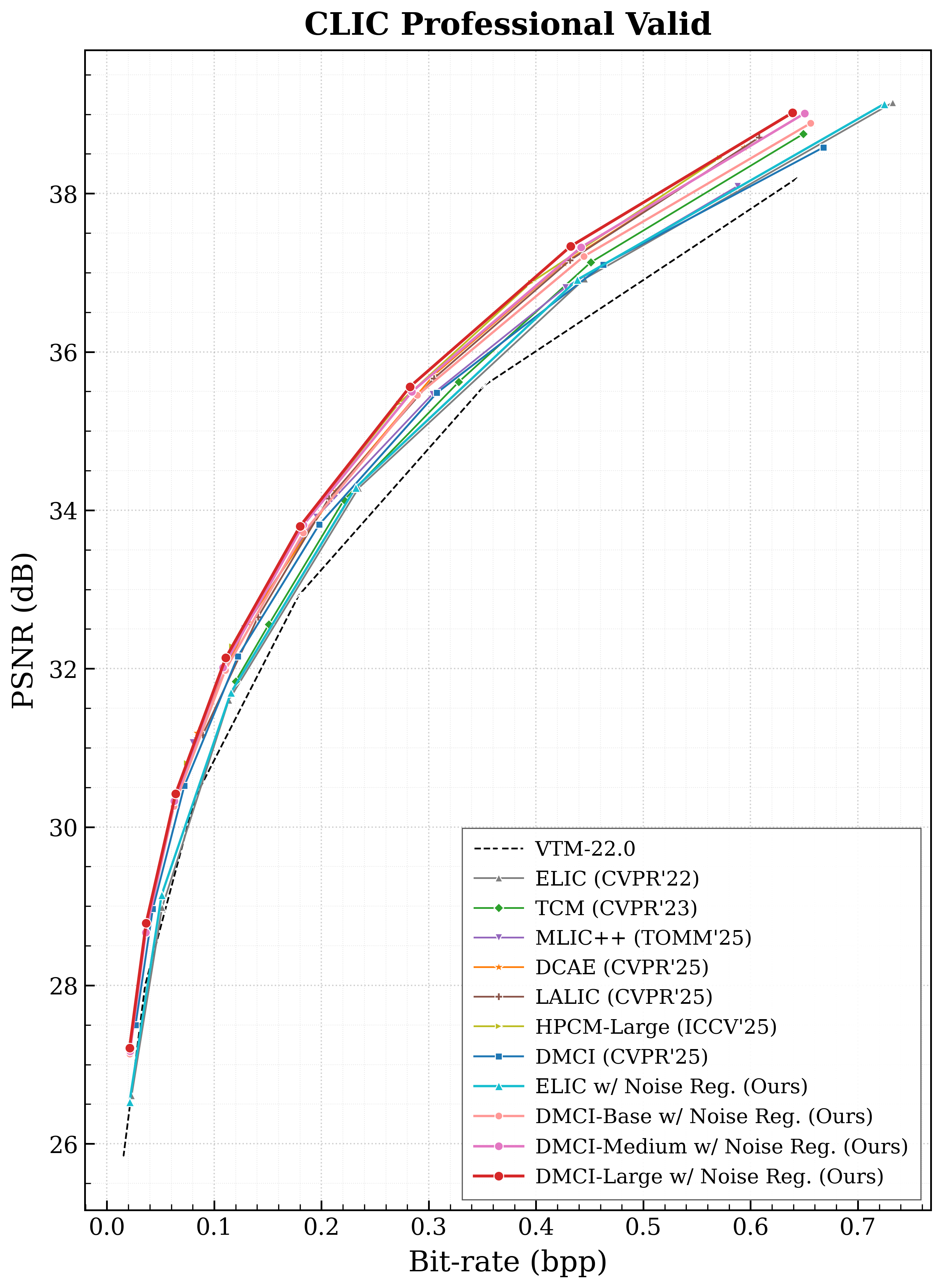}
  \caption{RD curves on \textbf{CLIC Professional Valid}.
           }
  \label{fig:rd_clic}
\end{figure}

\begin{figure}[p]
  \centering
  \includegraphics[width=\linewidth]{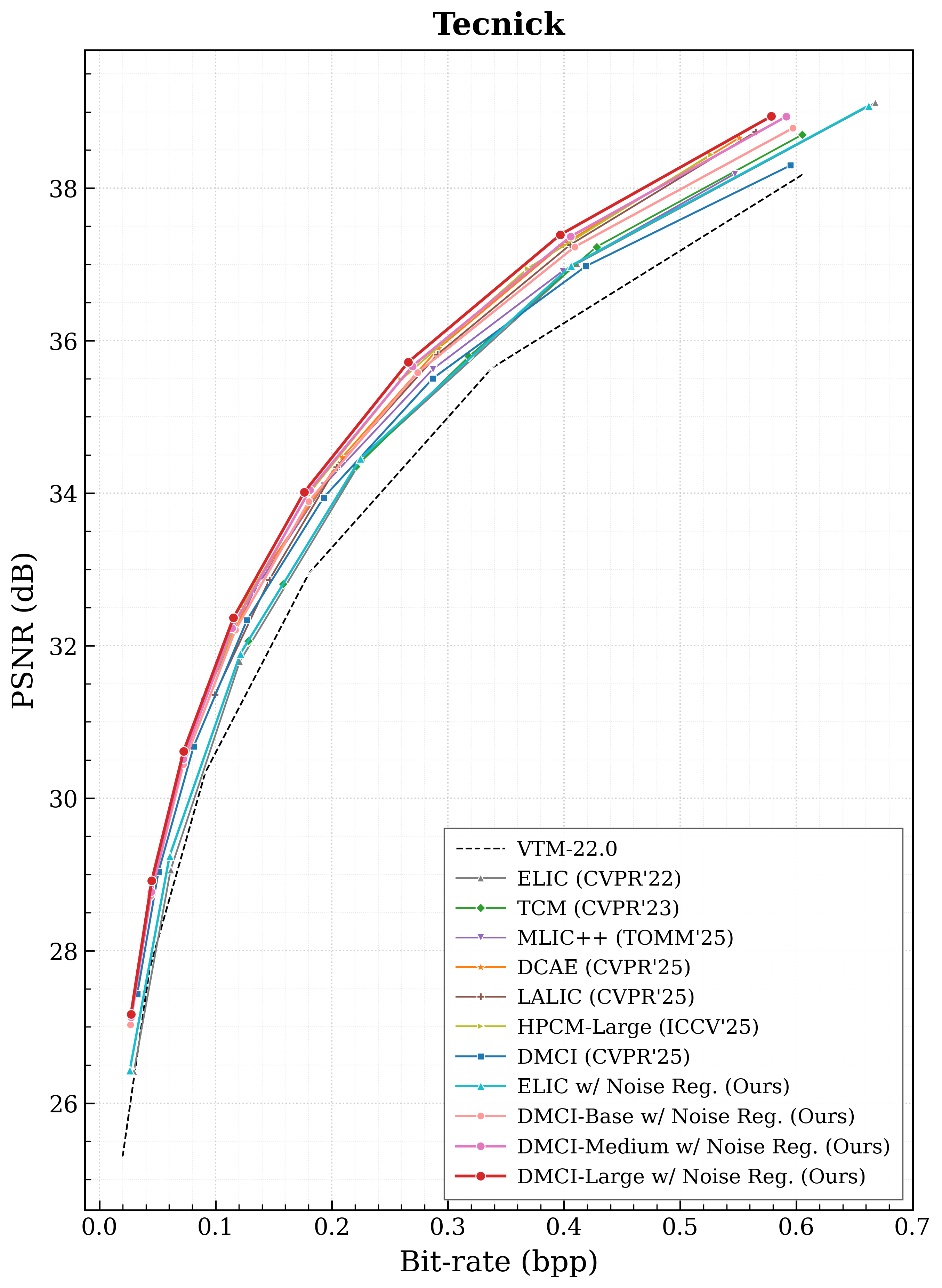}
  \caption{RD curves on \textbf{Tecnick}.
           }
  \label{fig:rd_tecnick}
\end{figure}

As a visual complement to the BD-Rate numbers in
Tables~\ref{tab:bdrate_results} and~\ref{tab:vtm_main},
Figures~\ref{fig:rd_kodak}--\ref{fig:rd_tecnick} plot the full
PSNR-vs.-bpp curves on Kodak, CLIC Professional Validation,
and Tecnick.

\end{document}